\documentclass[10pt]{article}
\usepackage[explicit]{titlesec}
\usepackage{hyphenat}
\usepackage{ragged2e}
\RaggedRight

\usepackage{algorithmic}
\usepackage{multicol}

\usepackage{rotating}  
\usepackage{theorem}   
\usepackage{bm}  
\usepackage{amsmath,amsfonts,amssymb}
\usepackage{booktabs}  
\usepackage{pdfsync}   
\usepackage{graphicx}
\usepackage{latexsym}
\usepackage{amsmath}
\usepackage{amssymb}
\usepackage{fancyvrb}
\usepackage{subfigure}
\usepackage{colortbl}
\usepackage{enumerate}
\usepackage{pifont}
\usepackage{stmaryrd}
\usepackage{textcomp}
\usepackage{fncylab}
\usepackage{multirow}
\usepackage{paralist}
\usepackage{wrapfig}
\usepackage{longtable}
\usepackage[]{algorithm2e}
\usepackage[table]{xcolor}
\usepackage{lscape}
\usepackage{lipsum}

\usepackage{garamondx}
\usepackage[T1]{fontenc}
\usepackage{amsmath}
\usepackage{graphicx}
\usepackage{xcolor}

\usepackage{geometry}
\usepackage{fancyhdr}
\usepackage[labelfont={footnotesize,bf} , textfont=footnotesize]{caption}
\makeatletter
\renewcommand\tagform@[1]{\maketag@@@ {\ignorespaces {\footnotesize{\textbf{Equation}}} #1.\unskip \@@italiccorr }}
\makeatother
\titleformat{\section}
  {\normalfont}{\thesection}{1em}{\MakeUppercase{\textbf{#1}}}
\titlespacing\section{0pt}{0pt}{-10pt}
\titleformat{\subsection}
  {\normalfont}{\thesubsection}{1em}{\textit{#1}}
\titlespacing\subsection{0pt}{0pt}{-8pt}

\makeatletter
\newcommand\sixteen{\@setfontsize\sixteen{17pt}{6}}
\renewcommand{\maketitle}{\bgroup\setlength{\parindent}{0pt}
\begin{flushleft}
\sixteen\bfseries \@title
\medskip
\end{flushleft}
\textit{\@author}
\egroup}
\makeatother

\title{Intelligent Vector-based Customer Segmentation in the Banking Industry}

\author{
Salman Mousaeirad*$^{a}$ \\ \medskip
$^{a}$Macquarie University, Sydney, Australia \\  \medskip
salman.mousaeirad@students.mq.edu.au
}

\begin{document}

\vspace*{.01 in}
\maketitle
\vspace{.12 in}

\section*{abstract}
Customer Segmentation is the process of dividing customers into groups based on common characteristics. An intelligent
Customer Segmentation will not only enable an organization to effectively allocate marketing resources (e.g.,
Recommender Systems in the Banking sector) but also it will enable identifying the customer cohorts that are most likely
to benefit from a specific policy (e.g., to discover diverse patient groups in the Health sector). While there has been a
significant improvement in approaches to Customer Segmentation, the main challenge remains to be the understanding
of the reasons behind the segmentation need. This task is challenging as it is subjective and depends on the goal of segmentation as well as the analyst's
perspective. To address this challenge, in this paper, we present an intelligent vector-based customer segmentation
approach. The proposed approach will leverage feature engineering to enable analysts to identify important features
(from a pool of features such as demographics, geography, psychographics, behavioral, and more) and feed them into a
neural embedding framework named Customer2Vec. The Customer2Vec combines the neural network classification and clustering methods as supervised and unsupervised learning techniques to embed the customer vector. We adopt a typical scenario in the Banking Sector to highlight how Customer2Vec significantly improves the quality of the segmentation and detecting customer similarities.

\section*{keywords}
Business Process Analytics; Data Curation; Customer Segmentation

\vspace{.12 in}


\section{introduction}

In this section, we present the outline and overview of this paper. Then, we explain the problem statement which is understanding the goal of the analyst in a supervised customer segmentation model. We also describe a motivating scenario in the Banking Sector. Finally, we present our contribution to highlight how the proposed method significantly improves the quality of the segmentation.

\subsection{Overview}

Customer segmentation is a significant concept in the road-map planning of organizations and plays a noticeable role in marketing strategies. An effective customer segmentation results in efficient marketing resource allocation as well as maximising the profit by offering the right products and services to the right group of customers and raising cross and up-selling opportunities~\cite{E2016AColombia}\cite{DBLP:journals/fgcs/SerhaniKSNBB20}.

One of the main challenges of customer segmentation is the identification of customers' features that are used to group the customers based on the similarity of these features. There are various features that are used in customer segmentation which are selected from a pool of features such as demographics, geographic, psychographic, behavioral, and more according to the goal of customer segmentation. Demographic features contain attributes such as age, gender, family size, income, and education. Geographical features include country, region, city size, and climate. Personality traits and benefits are examples of psychographic and socio-cultural features respectively~\cite{E2016AColombia}. Identifying and selecting important features (from a pool of features) is challenging as it is subjective and depends on the analyst's perspective. Moreover, the resulted segments are highly dependent on the features that are selected for the customer segmentation. On the other hand, the datasets of organisations do not necessarily contain all of these features. Therefore, feature engineering, i.e., the process of using domain knowledge to extract features from raw data via data mining techniques, and using relational datasets and integrating them to have all the required features for the customers is also very important~\cite{SimonsonEricJain2014AnalyticsREPORT}\cite{DBLP:conf/edbt/BeheshtiBM16}.

One of the main challenges of customer segmentation is understanding the goal and provide the segmentation based on this goal. For a general and non-goal-based customer segmentation, all customers in the same segment will receive the same offers, services, and policies~\cite{Wang2018ApplicationSystem}. However, from an analyst's point of view, it is likely that while two customers are in the same segment due to a specific attribute, at the same time they are in different segments according to another attribute. For instance, while they are in the same group in terms of product types they are using, they might be on different levels according to their level of loyalty or risk. Therefore a need for goal-based customer segmentation, i.e., customer segmentation based on a specific attribute or a combination of some attributes (according to the need of analyst) is one of the main challenges of conventional customer segmentation. In other words, instead of offering the same products and services and applying the same policies of different categories to the members of a group, for each attribute, we need a specific and subjective segmentation and offering based on this.

In this paper,  we present an intelligent vector-based customer segmentation approach which is a novel intelligent pipeline for goal-based subjective customer segmentation. This model is able to generate customer vectors and create segments according to one or a combination of a set of attributes. As an example in the banking domain, when the banking analyst plans to offer loan products to different customers, they first run the segmentation method based on the loan default risk of customers, and then, they can offer different loan products to the customers of different segments. The customer vector can be generated based on any customer attributes including usage, risk, credit, etc. Moreover, this model by generating a vector for each customer is able to calculate the level of similarity between customers. This will be helpful for the analysts of different areas such as baking, health, education, etc, to first find similar customers to a specific profile based on their needs and then, apply the required policies to them.

\subsection{Problem Statement}

A vast amount of research has been done in customer segmentation domain, yet the need for a goal-based and subjective segmentation method still exists. There are different customer segmentation approaches in academia and industry, including expert judgement, mathematical models, and big data analysis approaches such as text and sentiment analysis with the goal of grouping customers based on their behaviour similarities. These approaches are pursuing growth in the profit of organizations by applying similar offers, rules and policies to each segment (e.g., promotions, rival campaigns, large scale economic offers). However, one drawback of the conventional methods is providing the same segments for different goals. It is vital to design a method to segment the customers specifically for a goal defined by the analyst who is using the segmentation model. For instance, segmenting customers based on their purchasing behaviours, credits, etc.

To address the above challenges, in this paper we aim to develop an analytical hybrid model of customer segmentation. We present a general-purpose intelligent customer segmentation pipeline which consists of the supervised and unsupervised machine learning applications to extract a vector for each customer. This vector represents customers' behaviour in terms of one or a set of specific attributes based on analyst needs. In other words analysts of various domains such as health, finance and education will be able to apply this model when they need to segment customers based on the desired attribute. In this paper, we focus on customer segmentation based on one of the critical risks that banks are dealing with which is the credit risk. Credit risk refers to the potential loss of banks in case that their customer fails to pay their debt~\cite{RApostolikCDonohue2009FoundationsRegulation}. Thus, the generated customer vector displays the customer risk-based behaviour and by tracking this vector, we will create a dynamic map of customer pattern in terms of credit risk.

Using this vector the analyst will be able to measure the similarity between different customers according to the mentioned set of specific attributes. For instance, we can use this vector regarding the motivating scenario of segmenting customers based on their level of credit risk. When a bank detects a high-risk customer or when a customer is unsuccessful to re-pay the loan, the system will be able to measure the similarity of other loan applicants to this case. By this method, the loan application profiling system will improve in the identification of customers who are likely to default their loans.

\subsection{Motivating Scenario}

Business Processes are set of related activities in an organization to achieve a business goal~\cite{DBLP:books/sp/BeheshtiBSGMBGR16}\cite{DBLP:journals/spe/BeheshtiBM18}\cite{DBLP:conf/bpm/BeheshtiBNS11}.
Processes can be structured or adhoc. Adhoc processes are quite data-driven~\cite{DBLP:conf/icsoc/AmouzgarBGBYS18}\cite{DBLP:conf/adc/MaamarSBB15} and knowledge intensive~\cite{DBLP:conf/bpm/BeheshtiSGABYSC18}\cite{DBLP:conf/assri/ShahbazBNQPM18}\cite{DBLP:conf/icsoc/SchiliroBGABYSC18}.
Customer segmentation is an example of an adhoc process and can facilitate Personalization, Effective Acquisition and Retention, and intelligent marketing.
After the global financial crisis banks are investing much more on detecting the risks, controlling and managing them by applying new technologies such as machine learning, expanding profiling systems, and deep regulations~\cite{Leo2019MachineReview}. While banks are pursuing the most possible return by investing in various projects and lending money to businesses and individuals, it is crucial for banks to have a clear and comprehensive image and understanding of customer's behaviour.
One of the most critical parts of banks strategies is their loan approval policies and profiling systems. This deals directly with profit and loss of banks. While granting loans to well-credit customers will result is more profitability, failure in identifying high-risk customers and approving their high amount of loans might cause financial disasters for banks. Thus, financial institutions are investing in new technologies in the calculation of credit risk and prediction of loan-defaulters~\cite{Zakrzewska2007OnEvaluation}.

Loan repayment is one of the main challenges of banks and financial institutions. Current profiling systems are widely using decision tree models to identify whether or not the customer is qualified for the loan. However, there is a need to discover the risk level of customers by their features and find out the behavioral pattern of customers with high-risk levels and detect similar customers. The accurate applications of customer segmentation will strongly assist the banks to provide different segments with the appropriate loan offers and decrease the loan defaults.

For this motivating scenario, we aim to extract a vector for each customer that represents their behaviour in terms of a specific attribute (credit risk in this paper) and the capability of similarity detection amongst customers. Performing customer segmentation based on the credit risk of customers and identifying groups of them with similar behaviour in terms of credit risk will enhance in the loan application profiling system.

\subsection{Contribution}

In this paper, we are proposing a subjective and goal-based customer segmentation method which helps the analyst to create customer segments based on their goal. We extract customer features and by augmenting the primary data with the domain knowledge base and applying machine learning models on this semantic customer data, we propose an intelligent novel method to extract a vector for each customer which demonstrates their behaviour according to the goal of segmentation. This vector will provide the visualized segmentation and the analyst will have a view about the position of each customer compared to others. To do this segmentation, we are using personality trait scores of customers in addition to other features that are commonly used in the segmentation to find and consider the impact of customer personality in the resulted vectors and segments. Moreover, the calculation of customer similarity is an outcome of creating the customer vectors which makes it possible for the analyst to measure and consider the similarity of a customer to other customers based on the required attribute. For example in our motivating scenario and processing customer's loan applications, in addition to the current profiling systems, the analyst can use this model to measure the similarity of customer vector to customers who already have defaulted their loans and by this way, improve the loan application processing.

\subsection{Summary and Outline}

This section briefly discussed the overview, motivation, problem statement and contribution of this paper. The next sections are organised as follows:

\textbf{Section 2} reviews the researches on customer segmentation methods. We will review the state of the art works on clustering as an unsupervised segmentation approach and classification as supervised one and also combinations of them and also different evaluation methods. Moreover, we review the application of customer segmentation and data analytic techniques in the banking sector which is the motivating scenario of this research. We also discuss some works on personality detection from customer texts.

\textbf{Section 3} describes the method we used for customer segmentation this research in detail. The proposed pipeline includes data extraction, personality traits concatenation, standardization, classification networks, embedding vectors, and clustering the vectors. We will present details of the proposed pipeline and designed network in this section.

\textbf{Section 4} presents the outcomes of our method. We discuss the tools, techniques and packages that we used for implementing our method. We first discuss the feature engineering and extraction of the data in the banking sector as our motivating scenario. Then we provide the reader with the results of classification accuracy, clustering indexes, embedded vectors, and the visualized segmentation output.

\textbf{Section 5} concludes the research by a review of the results and how they achieved the project goals. We also briefly discussed the future extension area of this research which the author aims to go ahead with.

\section{Background and State-of-the-Art}

In this section, we will first discuss the key terms related to this research including customer segmentation, classification, and clustering. Then by analyzing the related works, we will discuss the researches on intelligent methods~\cite{DBLP:conf/wise/BeheshtiBSS19} and machine learning models in the current works in customer segmentation in the literature. They include different cases of using supervised and unsupervised customer segmentation. We will discuss different methods of clustering as unsupervised and classification as supervised customer segmentation. We will also discuss the application of machine learning in the banking sector. We review the works and researches about applying data analytics/mining techniques~\cite{DBLP:journals/computing/BeheshtiBVRMW17} in the banking industry.  According to the motivating scenario of this paper in segmenting customers based on their loan default risk, in this section, we will review the main sources of banking risks and will focus on one of the main critical risk types which is the credit risk. Finally, as we aim to utilize the personality traits of customers as part of their specifications in our methodology, we will review the works on personality detection by analysing the texts and applying sentiment analysis techniques.

An intelligent Customer Segmentation will not only enable an organization to effectively allocate marketing resources (e.g.,
Recommender Systems~\cite{DBLP:journals/algorithms/BeheshtiYMGGE20}\cite{DBLP:journals/access/YakhchiBGO020} in the Banking sector) but also it will enable identifying the customer cohorts that are most likely
to benefit from a specific policy (e.g., to discover diverse patient groups in the Health sector).

Thus, to review and discuss different aspects of this research in the literature and step into the related works in these areas, this section is divided into three sections. Section 2.1 will discuss the customer segmentation concept and different approaches to applying customer segmentation. This includes unsupervised techniques such as clustering as well as supervised models as classification. Section 2.2 will focus on the banking industry, different types of risk, and customer segmentation in this sector. Section 2.3 will review the works on personality detection from customer texts.

\subsection{Customer Segmentation}

Customer segmentation has always been at the heart of attention for businesses. Philip Kotler has defined a market segment as “a group of customers who share a similar set of needs and wants”. Kotler believes that customers can be segmented based on two major variables: descriptive characteristics (geographic, demographic, and psychographic) and behavioral considerations~\cite{Barat2009GlobalManagement}. The main segmentation variables in Kotler’s framework include geographic region, City size, age, family size and life-cycle, gender, income, occupation, education, social class, personality, loyalty status, and attitude toward the product~\cite{PhilipKotler2015MarketingManagement}.

Customer segmentation is the process of dividing customers into different and usually separated parts according to their attributes and features which aims to find classes with similar needs and values and therefore, make it possible to deliver specific services and products and marketing plans to each class~\cite{Ren2010CustomerMining}. This process can help to focus on target groups and with customizing services, increase in their satisfaction and therefore, rising retention and profitability rates and declining customer defection~\cite{Stone1996ManagingManagement}and providing those specific customers detected through the customer segmentation as beneficial customers based on their purchasing behaviours with special offers~\cite{Hsieh2004AnCustomers}. This process will also help the marketing and planning team with reducing the number of possible scenarios in providing services and evaluating the gain in large datasets of a huge number of customers~\cite{Murray2017MarketSet}.

In today’s business environment, precise customer segmentation gives critical knowledge to companies to survive in the competitive market. One of the main competition areas is applying innovative intelligent approaches to maximize the profit and minimize the loss. Demand forecasting based on accurate customer segmentation is an inexpensive way to achieve desired targets, as this leads to better customer experience, reducing lost opportunities, and efficient planning~\cite{Doganis2006TimeComputing}. Therefore, the significance of having a thorough understanding of the customer is emerging. In other words, while in the past most of the retailers based their products and services on the average of customers, in the modern and competitive market, product and service providers concentrating on providing different customer segments with customized products and services which will reduce their future business costs and will raise their profit~\cite{DBLP:conf/edoc/JalayerKBPM20}. They use different variables of customers to segment the customers depend on their products, services, and target customers. The most common variables for customer segmentation are below~\cite{Fisk2019AnSegmentation}:

\begin{itemize}
    \item Geographic variables including region and location information such as population data, density, and climate
    \item Demographic variables including age, gender, occupation, income, education, etc.
    \item Behavioural variables such as purchasing and consumption bahaviour.
    \item Lifestyle variables such as cultural practices, religion, and social values.
\end{itemize}

One category of customer features that plays an important role in this regard is behavioural and cognitive features and specific personality traits of customers. Moreover, one of the main sources to extract the customer’s personality is social media. Generally, with wide usage of social media by various people and businesses and growing trends of new services available through social media, it is clear that in the future even more valid data can be extracted from analyzing these valuable sources. On the other hand, the amount of available data is dramatically increasing, and therefore, businesses and government look at these sources as a precious asset and are willing to invest in data analytics in this area~\cite{Beheshti2019DataSynapse:Foundry}\cite{DBLP:conf/www/BeheshtiTBN17}.

The applications of customer segmentation utilize different methods however all of them are extracting and storing customer’s data, detecting the similarities and dissimilarities with using different models, and finally categorizing them into different homogeneous groups. Some works are using priori method to categorize customers which is extremely dependent on the level of expertise of the analyst who is performing this method. In contrast, most researchers are using standard key attributes of the customer such as demographic and socio-economic features including age, gender, income level, occupation, and education level~\cite{Murray2017MarketSet}.

There are different techniques used in the literature for customer segmentation which can be categorized as below~\cite{Sari2016ReviewEcommerce}:

\begin{itemize}
    \item Simple techniques which are using business rules, database query, and statistical data.
    \item RFM techniques which are segmenting based on recency, frequency, and monetary variables of customers
    \item Unsupervised segmentation, i.r., the clustering methods which aim to put customers with similar features in the same segment.
    \item Supervised segmentation which is applying classification techniques to segment customers based on labeling them according to a specific attribute.
\end{itemize}

Companies use data mining techniques as a guide to marketing strategies. Specifically, they are using machine learning applications in the process of customer segmentation~\cite{Dullaghan2017IntegrationCustomers}. The clustering technique is widely used in different implementations of market segmentation models and using different features and variables of data and bundling them into separate groups, plays a major role in the data mining process and therefore in market forecasting and planning research~\cite{Kashwan2013CustomerTechniques}.

To segment the customers into distinguished classes, different models are applied in the literature. Unsupervised learning has been used in the majority of researches with different methods of clustering and evaluation of clusters. While there are many advantages in using the standard models of clustering in such a high dimensional area such as well-separated clusters, sometimes when we need to segment the customers based on a specific attribute, the unsupervised clustering models may not create the meaningful clusters. In such situations, different models of classifications and also embedding models are commonly being used. These models include deep-learning classifiers~\cite{DBLP:journals/cee/NiuXAPBA20}, Support Vector Machine model, Random Forrest, Logistic Regression, etc. These models are evaluated in the literature with the methods of evaluating the results such as error rating, precision, recall, and F1 score~\cite{Ren2010CustomerMiningb}.

In the next subsections related works of clustering, classifications, and their evaluation methods will be discussed.
\subsubsection{Clustering}

Clustering is a vital data mining process and one of the unsupervised machine learning models frequently used in customer segmentation due to strategic and marketing plans. This is generally grouping the customers by detecting similarities between their behaviour and data which is able to automatically update the segments by receiving the present inputs of customers~\cite{Kashwan2013CustomerTechniques}.

Different methods of the clustering model have been utilized in the literature for customer segmentation in various domains such as health, banking, and education with large datasets and many features for customers. These multidimensional data might include some outliers as well. K-Means is by far the most common method in clustering, However, the K-Means classic model is not able to automatically recognize the number of clusters i.e., we need to give the number of clusters to the model as an input. There is also a random selection for initial centroids which might impact the accuracy of the model. The other drawback of k-means is being unable to detect the outliers of the dataset. Noise and outliers may deviate the clusters in the classic version~\cite{Zakrzewska2005ClusteringSegmentation}. However, there are some modifications of k-means in the literature that try to overcome the mentioned downsides by extracting outliers and finding the optimized number of clusters. Overall, this model in conjunction with its modified versions are the most common clustering method due to its successful performance in large and high-dimensional data ~\cite{Zakrzewska2005ClusteringSegmentation}\cite{KTHRoyalInstituteofTechnology2019CustomerAnalysis}\cite{Murray2017MarketSet}\cite{Namvar2010ASegmentation}\cite{Aryuni2018CustomerClustering}\cite{Chugh2020DataClustering}.

On the other hand, the model of Self Organizing Map (SOM)has the capability of distinguishing the number of clusters for the best efficacy as well as recognizing prominent features and suggesting cluster centroids however the performance is declining in working with a very large and complex dataset. There are some other clustering methods such as DBSCAN~\cite{Daszykowski2009Density-BasedMethods},Agglomerative hierarchal clustering\cite{Chugh2020DataClustering},Two-phase clustering algorithm for outlier detection~\cite{Jiang2001Two-phaseeDetection}, Ensemble clustering models~\cite{Cesario2016DistributedProblems},and K-Medoids~\cite{Aryuni2018CustomerClustering}and many authors utilized the mentioned clustering methods based on RFM scores. Some works used RFM scores directly for clustering~\cite{Noori2015AnSegmentation}\cite{Khajvand2011EstimatingContext}. Other works are available in the literature to evaluate the customer and make the customer segmentation by transforming the data into the RFM model and therefore, their performance is deeply dependent on the availability of the data which is making the RFM variables~\cite{Chao2008IdentifyingApproach}.In the next paragraphs, there is a review of some of the works that have been done on clustering in the literature. The methods and concepts will be described briefly.

Danuta Zakrzewska and Jan Murlewski have worked on clustering on a noisy and high-dimensional database in the banking industry. They applied three different clustering methods and compared the results. The first method was DBSCAN which is well designed to find noise inputs and works based on the concepts of density-reachability and density-connectivity and put the density-connected points in a cluster and detect the standalone points as noises. The second model they used was the classic k-means which is based on the iterative process of selecting cluster centroids as the average of clusters and assigning all points to the nearest cluster centroid. The third clustering method proposed by them was a 2 phase clustering algorithm for outlier detection. In this model in the first phase, they were doing a modified k-means by defining a certain distance that if a point’s distance to all clusters is more than this amount, then it will make a new cluster instead of joining to the nearest cluster in the classic format of k-means. Then in the second phase, the nearest clusters join together and any cluster with less than a specific number of objects is detected as noise. They used banking data sets and tried different experiments from the aspects of data size, dimensionality, and noise and compared the performance of three models. The results proved that k-means is working fine with a large amount of data with so many features however it is not successful in the case of having many outliers and results in deviated clusters. In contrast, DBSCAN recognizes noise but the performance has a high dependency on setting the model initial parameters which may result in poor outcomes and finding many noises if the mentioned variables are not suited for the case. On the other hand, the two-phase clustering method has a very successful performance in detecting noise. This is a really important attribute in the banking customer’s data as there are usually a lot of outliers in different customer feature data. The performance of this method will decline when the number of data dimensions increases~\cite{Zakrzewska2005ClusteringSegmentation}.

There are also some methods of clustering with a statistical basis in the literature. Sebastian Bergstrom discussed some of such methods. One of the methods is a modification of k-means. This model in the literature is known as k-means++. As one of the drawbacks of k-means is choosing initial cluster centroids randomly and therefore there is a risk of getting stuck in a local optimum, in the method of k-means++, they use a statistical algorithm to find cluster centers by a pre-defined distance from the first cluster centroid which is selected randomly. He also worked on the Mixture method based on mixed density distributions of the dataset with different clusters and finding the probability of each cluster with the mixture parameters. Gaussian Mixture Model (MCLUST), for example, is the mixture model with applying Gaussian distribution~\cite{KTHRoyalInstituteofTechnology2019CustomerAnalysis}.

They are some clustering works in the literature based on the RFM model (Recency, Frequency, Monetary). For instance, in ~\cite{Noori2015AnSegmentation}authors performed their clustering for mobile banking customers by using the score for each customer by simply multiplying R, F, M, and D scores (R*F*M*D). Here in this score calculation, R is the time from the last transaction (how many days), F is the frequency of transactions (how many times in a fiscal year), M is the average of the transaction amount, and D is the average of deposit of the customer. They set some thresholds for these scores based on expert advice and divided customers into some segments~\cite{Noori2015AnSegmentation}. In other researches, RFM attributes, as well as demographic features are used for clustering the customers by models such as SOM which is able to intelligently find the weight of each attribute and the number of segments in the efficient case~\cite{Hsieh2004AnCustomers}. Namvar et al. proposed a 2 phase segmentation model. They aim to use RFM and demographic attributes as the variable features of the k-means method~\cite{DBLP:journals/dase/TabebordbarBBB20}. They applied SOM on demographic features to find out the weight of features and as the result, education level, occupation, and age were selected as the most important features. Then, they implemented the k-means method by RFM attributes and divided the data set into 3 sections and for each section, they applied k-means this time by 3 mentioned demographic attributes (education, carrier, and age) which had the most influence and divided eacFh section into another 3 parts. Finally, for all clusters, they calculated LTV (lifetime value) and compared the results. RFM and LTV attributes together with demographic features are popular features being used in customer segmentation and clustering methods~\cite{Namvar2010ASegmentation}. Customer lifetime value will predict the future value of a customer by calculating and analyzing past and current values~\cite{Khajvand2011EstimatingContext}.

Murray et al. worked on different similarity detectors in their clustering research. While Euclidean straight-line distance is very common in the literature and different clustering methods, its performance is not accurate enough in the time series data which might include the outliers. They applied some outlier less-sensitive similarity methods such as the cross-correlation distance method (CCOR) and the Dynamic Time Wrapping method (DTW) which resulted in better outcomes. To perform clustering, they applied multiple models. Regarding k-means, the outcome and accuracy were dependent on the number of clusters that must be assigned before starting the model. They applied the SOM model as well which is an Artificial neural network~\cite{DBLP:conf/ijcnn/KhatamiNB0NZ20}\cite{DBLP:conf/intellisys/SchiliroBM20} and while the advantage of this model is in finding the efficient cluster numbers and centroids, when the dimension of the data (number of features for each customer) is very high, running the model will be computationally expensive. So, a data compression algorithm such as PCA is needed before applying the model~\cite{KTHRoyalInstituteofTechnology2019CustomerAnalysis}. The third method they applied was the hierarchical clustering method which works based on the distance matrix of customers and makes a bottom-up hierarchy method. This model also doesn't need the number of clusters to be assigned and it can find an efficient cluster number. One of the advantages of this method is the capability of this model to visualize the connection of clusters which helps in better model evaluation. By comparing the clustering methods on different datasets, they conclude that the best similarity detector depends on the type of data and computational capacity, evaluation, and the clustering method that are being used~\cite{Murray2017MarketSet}.

Saryu Chugh and Vanshita R Baweja proposed a two-phase customer segmentation model for online customers. They aim to overcome the weak performance of some clustering models in case of having noise in the dataset. Some models can’t detect the noise while some others recognize too many customers as noise and are not clustering and processing them. To address this challenge, they applied a 2 phase method with customer's features including, age, expenditure in online shopping, income, etc. In the first phase, they used a modified k-means in which before assigning the customers to a specific cluster, they calculate their distance to all clusters and compare it with the distance of the 2 closest clusters. If the distance of a customer feature vector to the nearest cluster is more than the minimum distance of 2 cluster centers, then this customer will create a new cluster itself. In the second phase, the agglomerative clustering is designed to merge the nearest clusters. This model presents a good performance in the detection of outliers and segmentation of existing customers and new ones~\cite{Chugh2020DataClustering}.

Ren et al. also worked on the banking dataset and proposed a 2 stage method using fully connected neural networks in the first phase and k-means in the second stage. They used behavioral and demographic features as well as monetary attributes such as the amount of loans. They performed the pre-processing stage including cleansing and filtering against their data. As the data is very huge with multiple attributes of socio-economic features, personal features, and transactional attributes, there will be a vector for each customer with their features of different categories. By applying SOM in the first phase and using an artificial neural network method, they find the number of clusters and the cluster centers. Then they use these outputs of SOM as the input of k-means in the second phase and assigned each vector to a specific cluster and continued the repeated process of cluster centroids and assigning each vector to the closest cluster until the steady state~\cite{Ren2010CustomerMining}.

Nan-Chen Hsieh also worked on customer segmentation on a dataset of banking customers in Taiwan. They used the self-organizing map method (SOM) for clustering the data. In this study, they used purchasing and repayment behaviours as attributes to identify customers with similar patterns in the same group in the Taiwanese credit card user database. Repayment Ability is defined as the “proportion of the number of months without delayed pay off over the number of months of holding the credit card"~\cite{Hsieh2004AnCustomers} and RFM scores represented purchasing behaviours. After distinguishing the clusters by applying SOM, they investigated on each group by demographic and geographic specifications and detected the most profitable group of customers~\cite{Hsieh2004AnCustomers}.

\vspace{-0.04cm}

Aryuni et al. worked on a clustering model of banking customers according to the data of their internet banking in a bank in Indonesia. Amongst the attributes of internet banking they used transaction date, transaction amount, customer ID, and account number and developed the RFM model. For clustering, they applied two models of k-means and k-medoids and compared the results. The logic behind the k-medoids method is to “minimize the sum of dissimilarities between data points within a cluster”~\cite{Aryuni2018CustomerClustering} and it checked every point within the cluster to check the increase or decrease in the cost in case the cluster medoids replace with this point. They applied both k-means and k-medoids for different values of k and used AWC and DBI methods for evaluating the clustering results and finding the errors. The research found the most optimum number of clusters as 2 and also better performance for the k-means method compared to k-medoids were concluded~\cite{Aryuni2018CustomerClustering}.

Taeho Hong and Eunmi King also worked on segmenting the customers of online stores. They used different features for their work and performed their clustering based on their psychographic features such as customer’s attitudes and interests. They proposed a 3 phase method. In the first phase by using the SEM model they aimed to find the features which have the most impact on the customer buying intention. Then in the second phase for clustering, they used both k-means and SOM methods to have more accurate results. If the outcomes of both methods for a customer were the same group, they assigned the customer in that cluster otherwise if the outcomes were not the same, they assigned the customer to a new class. Finally, in the third phase by applying the k-nearest method they found similar customers in segments~\cite{Hong2012SegmentingPurchase}.

\subsubsection{Classification}

There are also some works within the literature for segmenting the customers with the classification approach which is labeling the customers based on a specific attribute including money laundry detection, churn prediction, credit score, default risk prediction, etc. Particularly in the case of lending money, the pioneer and successful financial institutions invest in a thorough classification of borrowers~\cite{Mihova2018ABanks}. In these researches mainly in the finance area, authors used supervised learning on a combination of transactional, socio-demographic, and behavioural specifications of customers and applied different methods of classification to the dataset to categorize their data into separated classes with a labeled attribute~\cite{Phan2019SegmentationVietnam}.
Depends on the data specifications such as the size of the dataset, complexity, number of attributes, and dimensions, different classification methods can be applied including neural networks, support vector machines, Random Forrest, logistic regression, etc. These classifiers need a large size of labeled data to be trained and truly detect the pattern. There are also using some unlabeled to frame the clusters from big data~\cite{Gandhi2017TheData}. On the other hand, due to maximizing the model prediction accuracy there are also some researches that multiple classifiers have been used and then with bagging or boosting method, the results are being integrated and by voting between the outputs, the customer class has been selected. Some of these works are reviewed in the next paragraphs with a brief overview of the methods applied and the result of the researches.

Priyanka S.Patil and Nagaraj V.Dharwadkar worked on the classification of banking customer’s data which is large and complex in terms of high dimensions. They used one dataset for fraud detection attribute classification and another one for customer retention prediction. Their datasets include socio-demographic features such as age, gender, education, job, and income as well as banking history information like credit card details, transaction details, account details, and loan type. After processing the data and convert them into numeric variables, they applied artificial neural networks. They trained the model with 70 percent of data and tested with 30 percent and got 98 percent accuracy which proved their assumption that ANN classifier has a good performance on large, high-dimensional, and complex datasets such as banking customers features~\cite{Patil2017Learning}.

Carneiro et al. worked on a classification research for a case of fraud detection. The authors applied 71 features for each customer including demographic features such as gender as well as banking attributes such as time since first order and payment type. They applied three machine learning methods: SVM, Regression, and Random Forrest with training all models and comparing the results on cross-validation data. As this is skewed data, they used precision and recall to evaluate the models and Random Forrest's performance was the best. They also extracted features with more importance. The most contribution was for “the time since first-order” attribute and features like order value, similarity, and payment attempts had high contribution~\cite{Carneiro2017AE-tail}.

He et al. performed their research with the goal of churn prediction of churn rate with a banking data set. They applied SVM and logistic regression and compared the result. They also used the F1 score to evaluate the models based on precision and recall. The research resulted that the SVM model outperformed logistic regression~\cite{He2014PredictionModel}.

Bhagyashri S. Gandhi and Leena A. Deshpande proposed the application of ensemble classification. They discussed two methods of ensemble classification. The first is the bagging method in which there are some different classifiers each is being trained by part of the training data. After training when the model wants to select the class for input data, all the classifiers will give their scores for each class, and then by assigning weight to them and a voting method the class with the maximal vote will be selected. In contrast, there is a Consensus ensemble model in which if all classifiers have the same result, then that class will be the output of the classifier otherwise, the model can’t classify the sample. Therefore, while the accuracy of the consensus method is higher, many of the samples won’t have any outcome but in contrast to the bagging model, even if different classifiers come to different results, the voting method will choose the class with the highest rate. The authors applied the methods with different datasets and compared the results with a single classifier. They conclude that the ensemble method can result in fewer variance outcomes and thus more accuracy. Besides, with this ensemble method, working with a very large high dimensional dataset and complex classification by breaking it into separated parts with fewer dimensions and features and classifying each and using the voting or consensus approach, will result in more accurate outcomes~\cite{Gandhi2017TheData}.

Danuta Zakrzewska also worked on a hybrid classification model for loan repayment which used decision tree classifiers for creating rules on the classes extracted from clustering. In this research, they used numeric attributes of customers including deposit, account, payment rate, income, age, and the number of years employed, and divided these features into two groups. The first group used to make a clustering and then for each cluster by using the second group of features and applying the ID3 decision tree which has a great performance in noisy and multidimensional datasets, create the rules with the goal of predicting the customers who are not able to repay their loans i.e, evaluate the credit for each of clusters individually. They applied the method on datasets of a German bank and a Japanese bank clients and the results proved that this integration of these 2 models enhanced the accuracy of loan default prediction compared to using clustering or decision tree separately~\cite{Zakrzewska2007OnEvaluation}.

\subsubsection{Evaluation of Segmentation Methods}

While traditional customer segmentation models were empirical and used priori method, it is not possible to assign a level of value to each segment especially for the large multidimensional datasets we have in the banking industry for the customers~\cite{Ren2010CustomerMining}.

Evaluating the segmentation has been done with different methods through the literature. The Cluster Value Index (CVI) might use an internal approach which is the level of similarity inside the clusters and how much the clusters are compact and separable from each other. A high-value clustering will have clusters with a high level of similarity inside the clusters and in contrast high dissimilarity among different clusters. On the other hand by the external approach evaluation of clusters is based on a specific pre-determined attribute to see how accurately each cluster can get a label regarding that given attribute~\cite{KTHRoyalInstituteofTechnology2019CustomerAnalysis}.

Most of the works on the literature utilized these internal criteria as their CVI. For example, Aryuni et al. applied two methods of internal approach for their evaluation of segmentation. One index is AWC (Average Within Cluster) distance which is the sum of the squared distance of each point within the cluster to the cluster center. The other index is DBI (Davies-Bouldin Index) which is a ratio to compare the distance inside a cluster with the distance between the cluster and other clusters. Both mentioned indices are minimized for a successful clustering~\cite{Aryuni2018CustomerClustering}.On the other hand, Namvar et al. used an external attribute (LTV) to evaluate the clustering performance in their research as their focus was mostly on finding the target customers with the maximum value compared to finding just a cluster including customers with similar features~\cite{Namvar2010ASegmentation}.

\subsection{Application of Customer Segmentation in Banking Sector}

In recent years, several new organizations have emerged in the financial and banking industry, which has led to increased rivalry between competitors to attract new customers and keep the old ones. Therefore, the need for accessing customer knowledge has dramatically increased regardless of the size of the business. Understanding each customer will help businesses to provide specific service and marketing programs for each of them. Customer segmentation can facilitate this understanding and have an immense impact on sales prediction as well as other benefits including customer loyalty and brand reputation~\cite{Ezenkwu2015ApplicationServices}.

Customer segmentation in the banking industry is performed due to various goals. The main goal is offering the right products to the right customer groups to maximize the profit~\cite{Yuping2020NewLearning}. Especially according to the existing competition between banks and financial institutions, this is critical for them to have a detailed understanding of their customers, perform useful segmentation, and provide customers of each segment with new and innovative products and services. Moreover, from the B2B point of view, a lot of retailers are using digital channels of banks to sell their products and services through these channels. Therefore, having a clear image of different groups of customers will assist banks in providing the retailers with the most appropriate segment of customers regarding their products and services~\cite{Yuping2020NewLearning}.

The application of customer segmentation is not limited to sales and marketing goals. There are some researches on novel methods of measuring customer risks by applying supervised customer segmentation models~\cite{NamvarCredit}. Moreover, having a clear understanding of the customers is crucial in regards to addressing regulatory requirements. For instance, regarding detection of transactions suspicious to money-laundering, the significance of the concept of "Knowing Your Customers" (KYC) is noticeable~\cite{Tonsuk2011ANTI-MoNEYManagement}. Due to the massive amount of data available for banks about their customers, one main challenge is applying data analytics techniques to derive useful insights and finding customer DNA. Customer DNA in the banking industry is considered as the profile of the customer including all the data and features related to the customer. This includes financial aspects such as lifetime value, the total amount of accepted deposits, type and amount of lent loans, the response rate to banking campaigns, mobile banking credits, financial records and risks, Transaction features such as destinations, frequencies, and other related scales and in the same time, non-monetary aspects such as demographic characteristics, time and location of transactions, social media activities, etc. need to be taken into consideration~\cite{Bosnjak2011CreditBanking}.

\subsubsection{Data Analytics in the Banking Industry}

Nowadays, due to the high level of competition in the financial market for maximizing profitability by acquiring and retaining customers and also significantly growing in generated data from various sources, banks and financial institutions define investing in data analytics as one of the most critical investments that their life is significantly depended to that. It is mentioned in the literature that the amount of data generated in the last few years is noticeably more than all the data collected and available from beginning~\cite{Srivastava2018SuitabilityProfitability}. That’s why the businesses all need valid and processed data for their strategic planning and need to invest in creating the appropriated database, acquiring and storing the data from various sources, pre-processing the data including noise detection and normalizing, integrating, analyzing, and driving the insights based on client need~\cite{States201512A1}. This client might be external customers as well as internal customers including strategic planning units, sales teams, etc. Therefore, the data must be prepared and customized following the end-user requirements.

The approaches of dealing with data are changing fundamentally according to the massive amount of data generated every second i.e., around one million searches on Google per second, and the traditional methods of data mining are being replaced with modern data streaming in order to get the new information and eliminate the outdated data. Discovering the patterns from the data stream is the main goal of applying modern tools and techniques in data mining. While conventional static datasets can be used repeatedly, in the data stream analytics, there are some considerations including dataset size, limited memory, and real-time feedback which needs the application of novel methods for getting beneficial outcomes rather than sticking in a high amount of data~\cite{Nguyen2015AClassification}.

Financial institutions in recent years are more exposed to this data bombard and need to adapt their processes and procedures. The need for data mining as extracting a meaningful pattern from the massive quantity of data available in cyberspace is motivating banks and financial institutions to adopt and invest in Big Data to make competitive advantage in getting accurate data of their customers in order to maximize their profitability by providing convenient service and optimizing the processes~\cite{Gandhi2017TheData}.

Srivastava et al. categorized bank most critical challenges “linked to customers, competition, fraud, and compliance” and by using different data sources of customers including their transactions and loans information, and by applying Big Data Analytics technics in Fraud detection, Risk management, and CRM activities, they are trying to increase their revenue in the way of satisfying customer’s continuous needs and maximizing the efficiency and profitability ~\cite{Srivastava2018SuitabilityProfitability}. Utkarsh Srivastava and Santosh Gopalkrishnan also discussed the main applications of Big Data Analytics in the banking industry as customer lifetime patterns, proofing systems, customer segmentation and offering, sales and marketing planning, and Anti-money-laundry activities. They proposed a method and applied sentiment analysis for referral management and sales forecasting~\cite{Srivastava2015ImpactBanks}.

On the other hand, banks and financial institutions are dealing with a huge amount of multidimensional data which includes noise as well~\cite{Zakrzewska2005ClusteringSegmentation}. They are delivering many of their services in the current competitive market in the online platform and their database is significantly enormous and also with a lot of features and dimensions for each observation ~\cite{Chugh2020DataClustering}. Several other work defined data mining as revealing hidden useful patterns and knowledge from a noisy, vague, and multidimensional database also called Knowledge Discovery in Database (KDD)~\cite{Ren2010CustomerMining}\cite{DBLP:conf/springsim/BeheshtiM07}.

The data stored and processed in the banks contains demographic features such as age, gender, occupation, income level, education, etc. It also involves details of transaction history, credit card details, transaction date and amount, active accounts, available funds, credit scores, and merchant details~\cite{States201512A1}. According to the increasing trend of using mobiles to do banking activities such as simple transactions, balance checking, bill payments, installment payments, etc., considering m-banking features in the banking customer segmentation is being perceived more than before ~\cite{Noori2015AnSegmentation}. These features might include usage frequency, average transaction amount, and new offer contribution. Due to the high amount of the data and lots of features with different weights, preprocessing of the data with the goal of normalizing, selecting effective features and noise detection is vital before analyzing the data. Gao et al. by using the mentioned attributes of transactions worked on a framework for anti-money-laundry research. They applied data analytics and data mining techniques to discover transaction patterns and recognize unusual transactions and detect outliers and noise in the dataset to discover frauds and financial crimes~\cite{Gao2007AResearch}.

One of the most important methods in data analytics and data mining for the banking large and multidimensional database in order to support marketing activities is customer segmentation based on the various banking and non-banking features in accordance with the knowledge base of the banking system. The initial applications of customer segmentation in the banking industry, utilized only the key features of customers such as age, gender, income, etc,. However, these applications did not have precise outcomes for banking goals. Later some expressive and behavioral features were added as well such as loan type, transaction details, merchant name, which needs to be transformed into standard numeric scores in order to make it possible to input the data into the model, perform statistical analysis and interpret the results~\cite{Kantardzic2011DataEdition}.
 This is assumed that the data mining results in some features which accurately demonstrate the customer patterns and behavior and in addition to that these features are existing within our customer database~\cite{Stone1996ManagingManagement}. While the more number of features might lead to a better data mining process and achieving more accurate patterns, it might cause a high computational expense and therefore this needs to be considered in the feature selection~\cite{Murray2017MarketSet}.

\subsubsection{Banking Risks}

\begin{figure}[t]
    \centering
    \includegraphics[width=0.7\textwidth, angle=0]{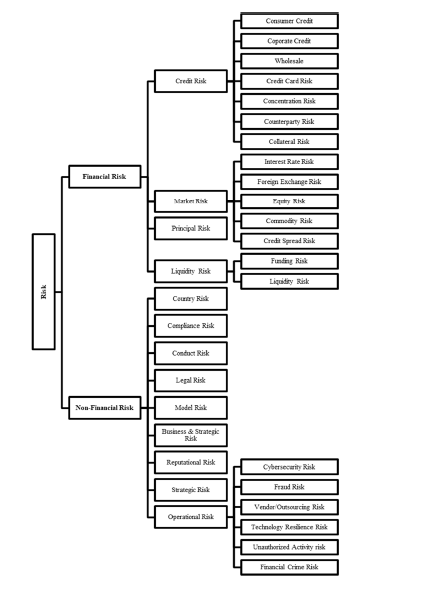}
    \caption{Taxonomy of banking risks~\cite{Leo2019MachineReview}.}
    \label{fig:Risk Sources in Banking Domain}
\end{figure}

Risk management has been one of the most crucial areas for banking and financial institutions over time however, its significance was manifolded especially after the global financial crisis. While having more detailed supervision from regulations on the one hand and the close competition between banks to satisfy customer segmentation, on the other hand, requires an investment in more accurate risk control policies, emerging new risk types, and storing huge amount of data in data warehouses encourages applying data analytics and machine learning models in the process of risk management. The main risk types that these financial institutions are dealing with can be categorized as Figure~\ref{fig:Risk Sources in Banking Domain}~\cite{Leo2019MachineReview}.

One of the main risk types i.e, the largest one that all financial service providers need to have an accurate prediction of that is credit risk which is defined as "the risk of potential loss to the bank if a borrower fails to meet
its obligations (interest and principal amounts)"~\cite{RApostolikCDonohue2009FoundationsRegulation}\cite{Leo2019MachineReview}.

There are several profiling methods that banks are using in order to predict the risk level of customers i.e., whether or not the customer will be able to pay back their loan on a regular basis. They are using a combination of customer data including account details, transaction information, income level, etc. Moreover, thanks to data analytics, there are some researches undertaken to segment the customers and predict risk levels in each segment~\cite{Nguyen2015AClassification}.

In this research, We are working on a novel customer segmentation in the banking industry with gathering and processing demographic and socio-economic features of the customers as well as transaction and banking history attributes. Then concatenation of personality attributes of customers based on the Big Five model to the processed and numeric vector of the mentioned attributes. Finally, applying both unsupervised and supervised learning methods with embedding models to classify and categorize different customer groups with different levels according to banking domain attributes such as risk, credit, etc.

\subsection{Detecting Personality Features From Customer Text}

There are some works in the literature regarding extracting personality traits by analyzing text. In such works in order to detect the personality type of a person, first one personality trait model is selected due to the target of this personality detection, and then usually by access to the texts of the person in social media such as tweets on Twitter and posts on Facebook, a sentiment analysis model that already trained with hundreds of thousands of texts will be used to score the personality traits.
Beheshti et al. proposed a data analytics pipeline to enable the analysis of patterns of behavioral disorders on social networks over time. They were using social media raw data such as tweets on Twitter or posts on Facebook and utilized the concept of knowledge lake~\cite{Beheshti2018CoreKG:Service}to automatically enrich and curate the data by summarizing~\cite{DBLP:journals/access/GhodratnamaBZS20}\cite{DBLP:journals/dpd/BeheshtiBM16}\cite{DBLP:conf/wise/BeheshtiBNA12}\cite{DBLP:journals/pvldb/HammoudRNBS15}, linking to knowledge lake and enriching the data and make it convenient for deriving characteristics such as personality traits. They also used LIWC as golden standards in the personality domain to analyze the curated texts to extract personality aspects~\cite{Tausczik2010TheMethods}. LIWC service provides more than 90 personality category scores for the texts. Besides LIWC, some available services such as Wikipedia and Google knowledge graph act as domain-specific knowledge base~\cite{Beheshti2019DataSynapse:Foundry}to contextualize the data by linking the tweet entities. Finally, they applied a neural network to map the input data with per feature as a dimension and link them to some behavioural disorders and by embedding technique, resulting in an embedded vector to represent the personality traits which can be used to find similar patterns of behavior and personality and also identify any dissonance in social media data~\cite{Beheshti2020Personality2Vec:Networks}.

Majumder et al. also had a work on personality detection from the text. They used the Big Five personality model and aim to find a sore for each of five personality traits including Extroversion, Neuroticism, Agreeableness, Conscientiousness, and Openness~\cite{Chiaburu2011TheMeta-analysis}. They trained 5 different networks for each of the personality traits and do a binary classification for each. They used the CNN model for finding feature extraction. Sentences of the text were vectorized by using word2vec vector embedding~\cite{Goldberg2014Word2vecMethod}and n-gram feature vectors. Then the vector was concatenated with some other features extracted from the text by applying LIWC. Finally, the input document vectors were connected to the 0 or 1 for any of the personality traits in 5 separated neural networks and the model was trained by the sentences of the document. Their test results proved that their method outperforms the state of the art methods~\cite{Ylmaz2020DeepTexts}.


\section{Methodology}

In this research, we are proposing a novel method of customer segmentation which is a combination of supervised and unsupervised learning models in order to generate a customer vector. This vector represents the behaviour of customers according to one or a set of specific attributes and can be used as an intelligent visualized tool that dynamically demonstrates customer position against other customers in a 3-dimensional coordinate system~\cite{DBLP:conf/www/BeheshtiTB20}\cite{DBLP:conf/wise/TabebordbarBB19}. This is also a useful approach to calculate the similarity of customers by using cosine similarity and distance similarity. By detecting similar customers, the analyst will be able to apply similar offers and policies to them. For instance, offering the services that are positively evaluated by a customer to similar customers as a marketing policy will be improved by applying this method. It also helps the analyst to prevent some frauds by detecting similar customer vectors to the vector of customer who has already committed a fraud. For instance, if a customer in the banking industry default a loan or is identified for a fraudulent transaction, the system will be able to immediately calculate the similarity between this customer vector to the vector of other customers and applicants to detect similar customers. As another example in the health industry, when a patient is detected in a special group of disease, it would be very useful to identify similar people who are exposed to the risk of having the same disease. We will discuss the pipeline and also details of the classification and clustering models used in our method in this section.

\subsection{Customer2Vec Model}

We are using a novel intelligent model including both supervised and unsupervised learning methods for customer segmentation based on one or a combination of some specific attributes. This is a general-purpose method of segmentation that can help analysts of various areas such as health, education, and banking in their goal-based customer segmentation. For example, in the banking sector, this segmentation can be credit-based or risk-based segmentation. Also in the health sector, we can segment customers based on their origin or disease type background. The overview of our proposed model for customer segmentation by embedding the customer vector is shown in Figure~\ref{fig:Customer2Vecmodel} which contains data pre-processing, data enrichment, feature extraction, and normalizing. The initial vector of customer features is made by concatenation of different types of features such as demographic, and behavioural. Once the initial customer profile is created, the next stage is classification and embedding the 3-dimensional customer vectors. In this stage, we fed the initial customer vector into a fully connected neural network with one hidden layer, and finally, we will cluster the embedded vectors into different segments based on one or a set of attributes based on the goal of customer segmentation. Different stages of the proposed model are as shown in Figure~\ref{fig:Customer2Vecmodel}:

\begin{figure}[t]
    \centering
    \includegraphics[width=0.8\textwidth, angle=0]{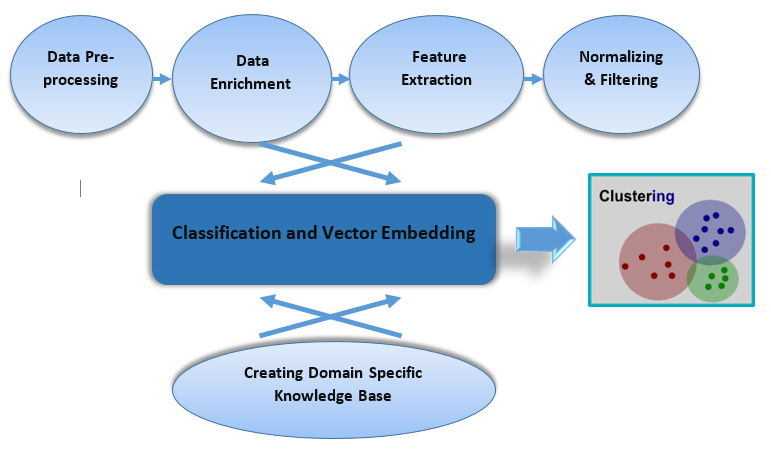}
    \caption{An overview of Customer2Vec Model.}
    \label{fig:Customer2Vecmodel}
\end{figure}

\begin{itemize}
\item	Data pre-processing: In this stage, we identify and extract the customer raw data which includes the features needed for the initial vector. Then by calculating statistical parameters such as mean and variance we clean the dataset, detect the outliers, and remove the invalid data. While it is likely that we don't have access to all required features for each observation, we will use intelligent techniques to replace these features with the average value of the same feature in other customers. We can also link different datasets based on common features. For example, link the data of customers with similar features such as age, gender, education level, and occupation.

\item   Data Enrichment: We need to enrich the raw data by using different knowledge-bases to make it ready for the extraction of features. For instance, any information about the location of customers can be enriched by the related information about this location. As an example in the health sector, by identifying the location of the customer, the background and historical records of disease in that location will enrich the data. Also in the banking industry, the location, source and destination of a transaction can be very important in concepts such as fraud detection or money-laundry.

\item	Feature extraction: The features that are used for customer segmentation are extracted from the enriched data. In a subjective customer segmentation, it is important to identify the features that have a high level of impact on the goal of customer segmentation. Also, it is dependent on the analyst's point of view to select the related features. We aim to use different categories of customer features including socio-economic, demographic, and behavioural features. For example, the customer segmentation in the banking industry, we are using banking and transactional features as well as demographic features. Moreover, we aim to concatenate the personality features of customers into their list of features~\cite{DBLP:journals/corr/abs-2001-04825}. For detecting customer personality scores, we use the fully connected neural networks to find the score of each personality trait of customers based on the Big Five model by using their texts on external datasets such as customer feedback and review. We concatenate these personality scores to the feature vector. The methods detect 5 binary values for each of the personality traits by analyzing the texts.

\item	Normalizing and filtering: We need to normalize our numeric elements of a customer's feature vector to ensure the cost function of the classification model converges to its minimum more efficiently. We also need to filter some special numeric features in a special range based on the domain analyst requirement. For instance, the analyst may be interested in the segmentation of customers in a specific range of income, age, etc.

\item Domain Knowledge-Base and linking data: We need to create a domain-specific knowledge-base to link our customer data to the external attributes of the context that we are doing segmentation~\cite{DBLP:conf/momm/GhafariJBPYO19}\cite{DBLP:conf/momm/BeheshtiHY19}\cite{DBLP:conf/wise/GhafariYBO18a}. This knowledge-base can be in education, banking, health, or any other area that we aim to implement the vector-based supervised customer segmentation. By creating this domain specific knowledge-base, we will be able to link the features of customers to the attributes of our domain. For example, if we aim to segment customers based on their level of risk in the banking sector, this domain specific knowledge-base includes any attributes related to a high-risk customer such as specific transactions (e.g.,gambling transactions), average transaction amount, age, and education. In this example, the gambling transactions of a customer will be linked to a high level of risk.

\item	Classification and vector embedding: Using a fully connected network, we link our initial vector with all different dimensions to a binary classifier which can classify them into 2 classes based on the domain specific attribute which is the goal of segmentation (e.g., credit risk in the banking sector or disease type in the health sector) and let the model learn from training data to find the optimized weights and then by using embedding method we will deliver the 3-dimensional vector which can be used to detect the similarity of customers by cosine similarity and distance similarity methods and find a group of similar customers. By this method of supervised customer segmentation, the analyst can offer specific services to customers with similar vectors which results in optimizing the profit. We will discuss the details of our classification and vector embedding network later in this section.

\item	Clustering the embedded vectors: Once we embed the customers' 3-dimensional vectors from the fully connected neural networks, we will apply clustering methods to these vectors to segment the customers. In this step, we are using four different clustering methods. These clustering methods are modified k-means, self-organized map (SOM), Gaussian mixture, and mean shift and will compare the clusters made by each method to find the most efficient one.

The classic form of the k-means model needs a pre-defined k (number of clusters) and selects the initial cluster centers randomly. However, if cluster centers are too close, the clustering cost function might be stuck in the local minimum which means that the clusters are not the most efficient possible clusters. To avoid this issue, we use a modified k-means with the below algorithm:

\begin{algorithm}
1.	Randomly select K cluster centers as K of our vectors and find their distance (for each pair of centers).

2.	Assign each point to the nearest cluster center.

3.	If the maximum distance of one point of one cluster to its cluster center is more than the minimum distance of this cluster center i from any other cluster center, go to 1.

4.	Find new centers as the average of points in each cluster.

5.	If any center changed, go to 2.
\end{algorithm}

We are using different clustering models and compare the total cost of them which is defined as the sum of the square distance between each point and its cluster center and choose the best model. The clustering models we apply are k-means, SOM, Gaussian mixture, and mean-shift.

\end{itemize}

\subsection{Classification and Embedding the Customer Vector}

We aim to embed a vector for each customer to provide the analyst with the capability of dynamic visualized segmentation. Therefore, in order to extract a 3-dimensional vector, we design the embedding layer with 3 units through our neural network, and then put another hidden layer and finally a binary classifier based on the attribute which is the goal of segmentation. For instance, if we are segmenting banking customers based on their loan default risk (which is our motivating scenario in this research), the classifier indicates whether the customer with a given feature vector in the training data already has had a loan default history or not. The goal of this example of classification is to predict the probability of loan default for a customer.

\begin{figure}[t]
    \centering
    \includegraphics[width=0.6\textwidth, angle=0]{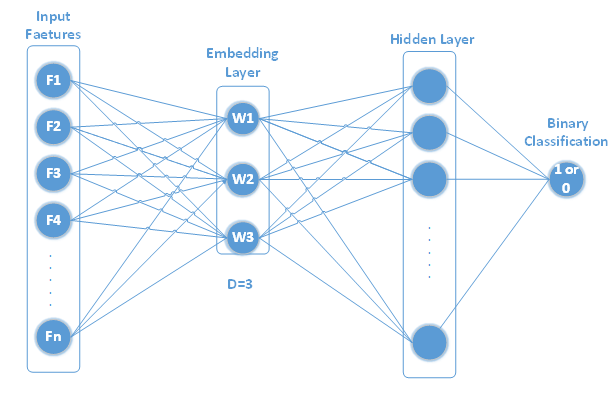}
    \caption{Vector embedding network extracts a 3-dimensional vector on its embedding layer.}
    \label{fig:Vector-embedding-network}
\end{figure}

The model uses the initial feature vector of the customer as input and by assigning the random weights in the neural network layers and forward propagation process, calculates the value of the end node as the probability of loan default risk. Then compares it to the actual value of this node, i.e., the loan default history of the customer, and calculates the error. Once the error is calculated, by using backward propagation the model will adjust the weights in all layers of the neural networks. The next iteration will do the same process with all training data by using adjusted weights. This process will continue until finding the optimal weights and minimum cost function. Finally, the model will create the 3-dimensional embedded vector for all customers in the 3 unit hidden layer by the product of the initial customer feature vector and the final weight matrix. Figure~\ref{fig:Vector-embedding-network} shows the classification and vector embedding network that we use in this model.

\vspace{-0.5cm}

\vspace{-0.5cm}

\subsubsection{Optimising the Cost Function}
\begin{figure}[t]
    \centering
    \includegraphics[width=0.7\textwidth, angle=0]{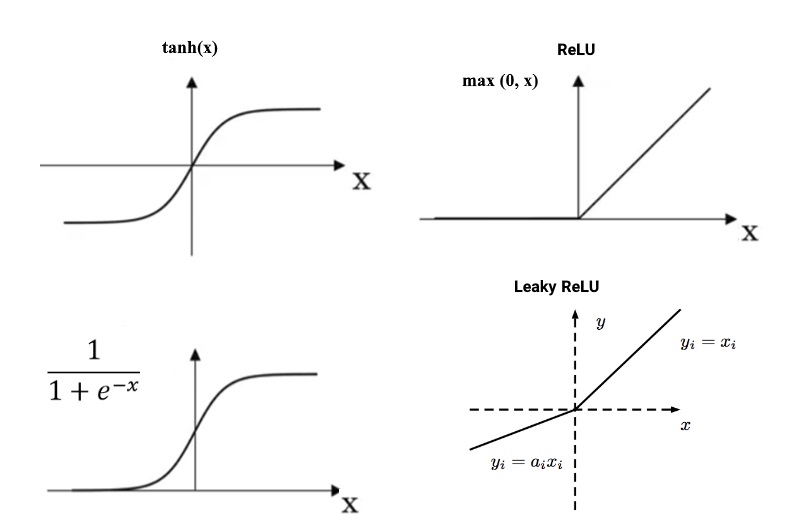}
    \caption{Activation functions used for vector embedding network.}
    \label{fig:activation functions.jpg}
\end{figure}

We are using 4 different activation functions and comparing the results and accuracy of the classification as shown in Figure~\ref{fig:activation functions.jpg}. Below are the mathematical forward propagation formulas that are used in the model:

  \begin{equation}
  \small
    h_ \theta (x) =  \frac{\mathrm{1} }{\mathrm{1} + e^- \theta^Tx }
  \end{equation}

  \vspace{-1cm}

  \begin{multicols}{2}
  \begin{equation}
  \small
    A^{(1)}=\theta^{(1)}*X
  \end{equation}\break
  \begin{equation}
  \small
    W^{(1)}=h(A^{(1)})
  \end{equation}
\end{multicols}

\vspace{-1.5cm}

\begin{multicols}{2}
  \begin{equation}
  \small
    A^{(2)}=\theta^{(2)}*W^{(1)}
  \end{equation}\break
  \begin{equation}
  \small
    W^{(2)}=h(A^{(2)})
  \end{equation}
\end{multicols}

\vspace{-1.5cm}

\begin{multicols}{2}
  \begin{equation}
  \small
    A^{(3)}=\theta^{(3)}*W^{(2)}
  \end{equation}\break
  \begin{equation}
  \small
    Classifier=W^{(3)}=h(A^{(3)})
  \end{equation}
\end{multicols}

In the above formula, $X$ refers to the initial vector of customers including their 63 dimensions of features which are standardized and each element of the vector is a number between 0 and 1. $theta^{(n)}$ identifies the weight matrix in the $nth$ layer which is initiated with random numbers and will be optimised in the process of forward and backward propagation to minimize the cost function. In the process of forward propagation, $A^{(n)}$ is the product of the output of layer $n-1$ by the weight matrix of this layer, and $W^{(n)}$ which is the output of layer n is calculated after applying the Sigmoid function on $A^{(n)}$.

After finding all the outputs in the forward propagation process, by calculating the difference between what is predicted by the classifier and the actual label of the output node, we will identify the difference and by backward propagation, we will use it to find out the derivative of weights for each layer. We need these derivatives for applying the optimisation process for our cost function.

\vspace{-0.5cm}

 \begin{multicols}{2}
\begin{equation}
    \delta^{classifier} =  W^{(3)}-y
  \end{equation}\
  \begin{equation}
    h^{'} (A^{(n)})=W^{(n)}*(1-(W^{(n)})
  \end{equation}
\end{multicols}

\vspace{-1.5cm}

  \begin{multicols}{2}
  \begin{equation}
   \delta^{layer2}=(\theta^{(3)})^T \delta^{classifier}*h^{'} (A^{(3)})
  \end{equation}\break
  \begin{equation}
    \delta^{Embedding layer}=(\theta^{(2)})^T \delta^{layer2}*h^{'} (A^{(2)})
  \end{equation}
\end{multicols}

The goal is to make the most accurate prediction by our binary classifier. Therefore we define the cost function as the binary cross-entropy function similar to the cost function of logistic regression to calculate the difference between $W^{(3)}$ and y for all the training data. Then by applying gradient descent, update the weight matrix in each iteration. Cost function and the partial derivatives using in gradient descent are coming from the below formula:

  \vspace{-0.5cm}

  \begin{equation}
    Cost(\theta)=\frac{1}{n}\sum_{i=1}^{n} \sum_{k=1}^{K}[y_k^{(i)}log((h_\theta(x^{(i)}))_k)+(1-y_k^{(i)})log(1-(h_\theta(x^{(i)}))_k)
  \end{equation}

  \vspace{-0.5cm}

  \begin{equation}
     \delta^{(n)}=\frac{\partial Cost}{\partial W^{(n)}}
  \end{equation}

  $n$ refers to the number of rows (observations) in our training data matrix and $k$ presents the number of classes in our classification which is a binary classification in our work in terms of whether or not the customer has had the loan default risk, however, this classification can be extended to multi classes in terms of the risk level.

  Finally, the goal is maximizing the Cosine similarity of embedded vectors for 2 customers with similar profiles. Therefore once the 3-dimension vectors are extracted from the model, the profiling system will be improved. Whenever the bank has a customer "t" with loan default, they will calculate the similarity of other customers such as "e" by calculating the Cosine of their vectors and detect the customers with more than a certain amount of similarity.

    \vspace{-0.5cm}
  \begin{equation}
\cos ({\bf t},{\bf e})= \frac{ \sum_{i=1}^{n}{{\bf t}_i{\bf e}_i} }{ \sqrt{\sum_{i=1}^{n}{({\bf t}_i)^2}} \sqrt{\sum_{i=1}^{n}{({\bf e}_i)^2}} }
\end{equation}

\subsubsection{Customer2Vec Versus Word2Vec}

\begin{figure}[t]
    \centering
    \includegraphics[width=0.8\textwidth, angle=0]{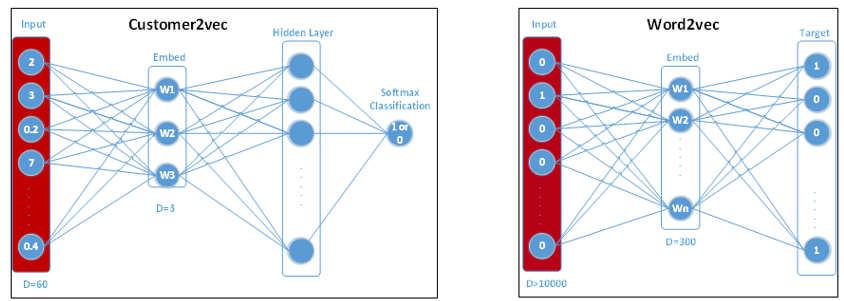}
    \caption{A comparison of customer2vec with word2vecindicates some differences in the input of the networks, dimension of vector, and how to link the objects in 2 models.}
    \label{fig:Customer2Vec-vs-Word2Vec-network}
\end{figure}

Embedding the customer vector in this model is using a similar network with word2vec. Word2vec is a widely used technique in sentiment analysis that assigns a vector to each word in a document. This vector represents the meaning of the word and the similarity amongst different words in the document. The word2vec model is also starting by assigning initial vectors to each word and then in each iteration based on the window size, the vectors will be optimised through linking words that are next to each other. Figure~\ref{fig:Customer2Vec-vs-Word2Vec-network} shows the structure of this model and compares it with the word2vec ~\cite{Rong2014Word2vecExplained}\cite{Goldberg2014Word2vecMethod} embedding model.

While the customer2vec model is acting similar to word2vec in terms of embedding a vector through a fully connected network, there are some aspects of the customer2vec model that distinguishes this model:

\begin{itemize}
\item	While word2vec is assigning one-hot vectors to each word at the beginning of training and then will find the optimised vector after all the iterations of training data, the customer2vec model is not using a one-hot vector for customers at the beginning of feeding the network but instead, we are using a normalized numeric vector including all real meaningful features of the customer. So we can expect better and quicker convergence in customer2vec.

\item	The embedded layer in the word2vec model has between 20 to 300 dimensions (depends on the size of the document), in customer2vec we add a hidden layer to make it possible to have a 3-dimension embedding layer. This will make it possible for the analyst to have a thorough visualization of customers after finding the vectors.

\item	The classification of the customer2vec model on the contrary of word2vec is not based on window size and linking customers to each other, but this model is using supervised learning. In other words, we have a label for each customer profile based on the required attribute. For example, if the goal is finding the customer vector based on their loan default risk, we label the training data whether or not they have defaulted any loan. Finally, the model can predict the probability of loan default for the new customer with a given feature vector.
\end{itemize}

\begin{figure}[t]
    \centering
    \includegraphics[width=0.4\textwidth, angle=0]{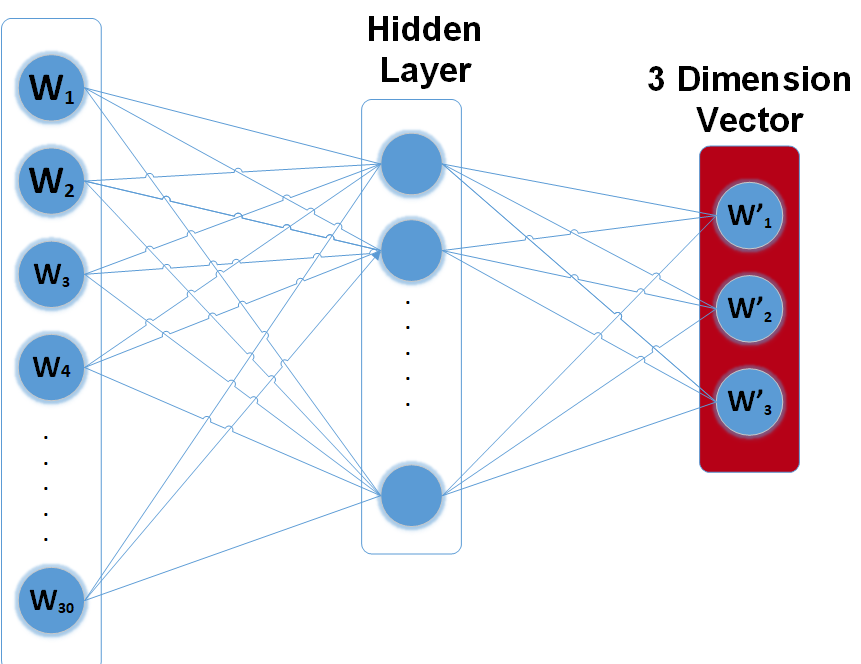}
    \caption{The network of embedding customer vector by data compression.}
    \label{fig:Embedding-customer-vector}
\end{figure}

We also use another network for extracting the 3-dimensional customer vector as and compare the accuracy of the classification model and extracted vectors with the main model to find the most efficient network for our goal. This model also uses a similar embedding network to the mentioned model but eliminating the 3-nodes hidden layer. We use a network with one hidden layer with 30 nodes and extract a vector with 30 dimensions and then by applying embedding models and data compression methods, by a fully connected network, the 3-dimension vector will be concluded as shown in Figure~\ref{fig:Embedding-customer-vector}.

\subsection{Clustering Embedded Vectors}
 We have embedded a 3-dimension vector for each customer and we can use the vectors to find similar customers and segment them. We aim to segment different customers in terms of different levels of related banking attributes which in this research is their level of risk and provide a visualized segmentation in the 3D coordinate system.
 In order to cluster the vectors, we are using the modified k-means, self-organized map (SOM), Gaussian mixture, and mean shift. We will compare the results of different method's clustering by comparing their accuracy indexes. The indexes we are using are Silhouette-score, Calinski and Harabasz score (CH), and Davies-Bouldin Index. We will use the clustering methods with different numbers of clusters and will find the most efficient number of clusters by knee method. However, in regards to the number of clusters, the goal of segmentation and the analyst needs also play an important role. For instance, if we apply this model for segmenting banking customers based on their risk level, the analyst might require to have 3 segments of low-risk, medium-risk, and high-risk customers.

\begin{figure}[t]
    \centering
    \includegraphics[width=0.5\textwidth, angle=0]{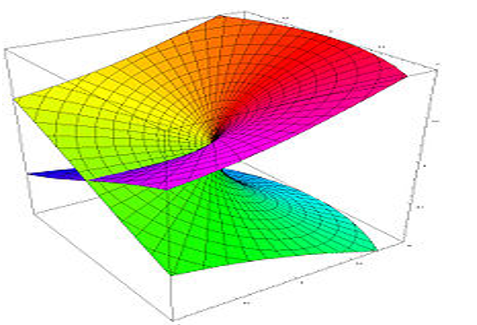}
    \caption{Vector-based segmentation provides the analyst with the visualized clusters in a 3-dimensional coordinate system.}
    \label{fig:Clustering-embeded-vector}
\end{figure}

 Figure~\ref{fig:Clustering-embeded-vector} shows a visualized vector-based customer segmentation. In this coordinate system, the customer vectors are divided into 5 segments and the zone of them are shown in different colors of purple, red, yellow, green, and blue. Each segment contains customers with similar vectors and the analyst will be able to assign a label to the segments by analyzing the profiles of customers of each segment. Moreover, by changing customer features over time (such as age or income), the position of each customer in the clusters may change. The analyst will be able to use this dynamic dashboard to identify the change in customer vectors and consider it in strategic and marketing policies for each customer.

\section{Experiment, Results, and Evaluation}

In this section, we discuss the results of applying our method to the banking customer data and the accuracy of our classification model as well as the evaluation of clusters and segments created by our model. We first present a banking motivating scenario and discuss our goal which is the vector-based customer segmentation based on the credit risk of the customers. We apply the Customer2Vec model to banking customers. Each observation, i.e., customer profile of the banking industry, contains banking and transnational features as well as non-banking features including socio-demographic features and personality traits. As mentioned in Section~3, we extract personality traits from customer texts based on the Big Five model.

We discuss our dataset specifications and discuss our intelligent classification network and the different activation functions, epochs, and batch sizes that we are using. We select the most efficient epoch and batch size by comparing the results of evaluations in the cross-validation data. Finally, we discuss the clustering of our embedded vectors with different clustering evaluators and provide the reader with the results and accuracy of our model with the various numbers of clusters. We use Python 3.7.7, Numpy 1.18.5, pandas 1.1.1, and Keras 2.4.3 for our coding.

\subsection{Motivating Scenario: Banking Domain}

In this paper, we apply the vector-based segmentation to the customers of the banking sector. We work on the customer features in the banking industry and extract the vector for each customer. As mentioned in the methodology section, the segmentation and resulted customer vectors are based on one or a set of attributes such as risk, money laundry, and credit. Therefore, we can provide customized customer segmentation for each attribute. In this research, we focus on the classification of customers to predict their loan default risk. We extract customer vectors and segment customers based on the customer's credit risk.

Our model utilizes all different customer features without pre-assigning any weight to them. The model discovers the level of their influence on the domain-specific banking attribute. We are dividing customer features into two groups of banking and non-banking features.

\subsubsection{Customer Banking Features}

Regarding banking features, we extract numeric customer features such as transactions, loans, credit cards, accounts, and m-banking activities. We are breaking each category of data to extract the detailed numeric data which is likely to have an impact on the level of customer credit risk. Features related to transactions, for example, are broken into average monthly value, frequency, transaction timing features (for example the proportion of transactions outside working hours), and specific transactions such as rents, mortgage, and gambling transactions. We use the domain experts' and banking managers' point of view about any feature that can be extracted which might have an impact on a customer's loan default risk. These banking features are available in the bank's internal and external datasets, and also public datasets. The main banking activity features are as below:

\begin{itemize}
    \item Loan including type, amount, default history, installments, and interest rate.
    \item Transactions including average monthly value, frequency, gambling transactions, and unusual transactions in terms of time or amount.
    \item Accounts including account balance, savings, term deposits, and available funds.
    \item Credit including credit score, credit card type, the average monthly value of the purchase by credit card, and any late payment.
    \item Engagement including loyalty plans, promotions, and cross-selling contribution.
    \item M-Baking including usage frequency, average value, and new offer contributions.
    \item Average number of visiting branch per month.
\end{itemize}

\subsection{Non-banking Features}
We are extracting non-banking numeric features of the customers such as demographic features. These features are also broken into detailed features such as age, AMR (Average Monthly Revenue), AME (Average Monthly Expense), education, and occupation.

In this research, we aim to concatenate the personality traits of customers into the list of customer features to enable considering personality features in the customer segmentation. We aim to find the weight and impact of different personality aspects to customer credit risk level in the banking sector which can improve the efficiency of customer profiling system for in loan applications of the banks. In other words, instead of using a decision tree of only banking historical data of customers to evaluate their credit and risk, we are working on a model to use banking, demographic, and personality features of them to get a better understanding of customers and their credit risk level.  One of the most common models of personality analysis is the Big Five model~\cite{Digman1990PersonalityModel}. In this model, the personality traits are categorized into 5 main attributes:

\begin{itemize}
\item	Extroversion (EXT): Being outgoing, talkative, and energetic versus reserved and solitary.

\item	Neuroticism (NEU): Being sensitive and nervous versus secure and confident.

\item	Agreeableness (AGR): Being trustworthy~\cite{DBLP:journals/access/GhafariBJPMYO20}, straightforward, generous, and modest versus unreliable, complicated, meager, and boastful.

\item	Conscientiousness (CON): Being efficient and organized versus sloppy and careless.

\item	Openness (OPN): Being inventive and curious versus dogmatic and cautious.
\end{itemize}

\begin{figure}[t]
    \centering
    \includegraphics[width=0.75\textwidth, angle=0]{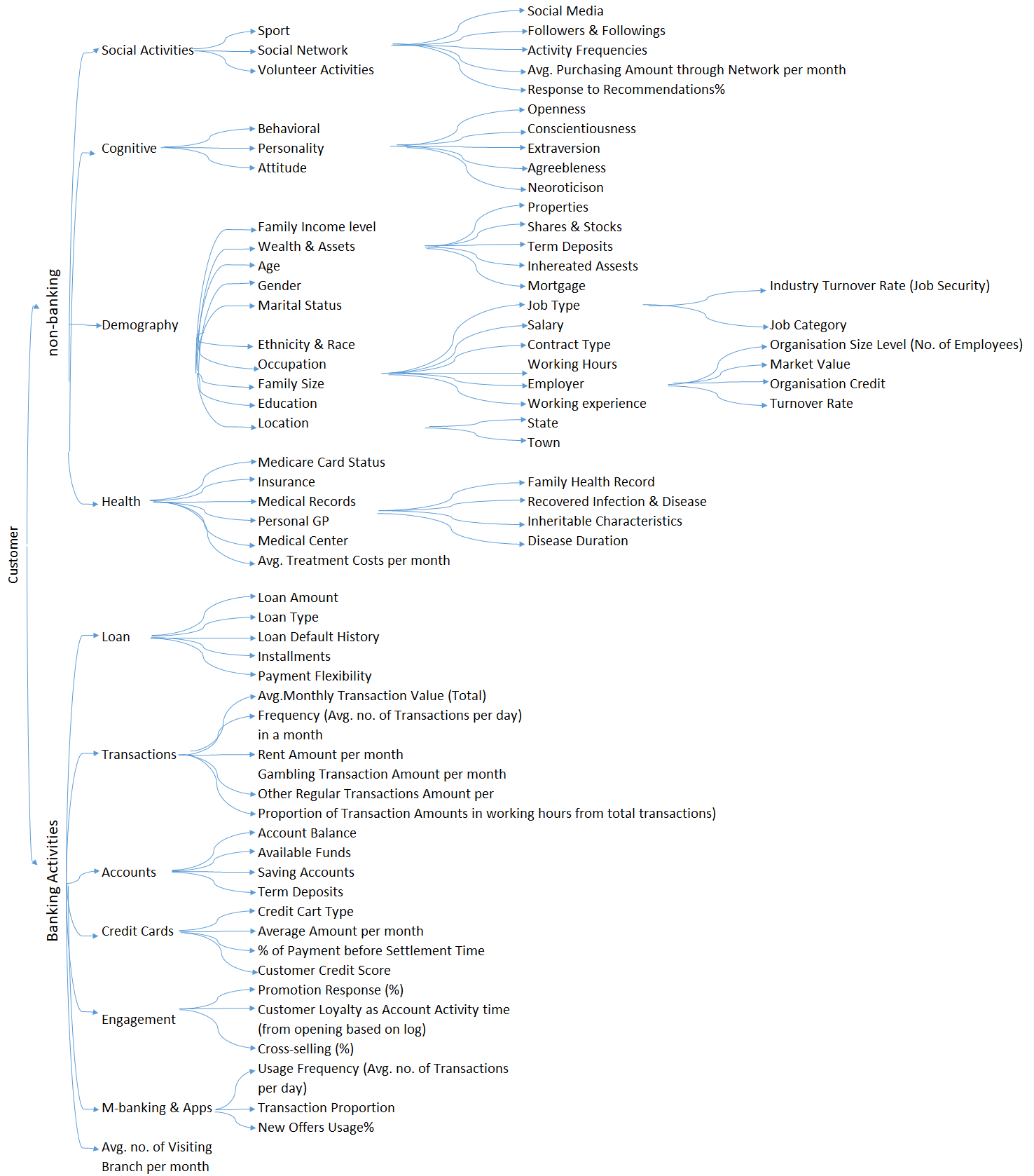}
    \caption{Customer Extracted Feature Breakdown.}
    \label{fig:Multilevel}
\end{figure}

We are using customer texts to extract customer Big Five personality traits. These texts are from customer's reviews and feedback in the external datasets or their texts on social media~\cite{Beheshti2020Personality2Vec:Networks} and are used for extraction of their personality trait scores. For extracting personality scores, we use an available service in GitHub for extracting personality scores~\cite{Ylmaz2020DeepTexts} \footnote{http://github.com/senticnet/personality-detection)}. By training a convolutional neural network with the texts of customers, this service can predict the personality traits of them as a binary classifier i.e, assign 0 or 1 to each of 5 personality traits. We concatenate the personality scores with all the banking activity and transnational features as well as demographic and socio-economic features to make the initial customer vector which is numeric for all features. This vector will be the input of the customer2vec network.
The feature breakdown of customers is shown in Figure~\ref{fig:Multilevel}.

While we extract all the customer data as numeric features, we will standardize them by removing means and dividing by the standard deviation. The final Excel or CSV file as our customer data lake is a table with observations as rows and features as columns (63 features for each observation in this research) and contains numeric scores for each feature. These scores are pre-processed and standardized as well.

\subsection{Dataset}

\begin{figure}[t]
    \centering
    \includegraphics[width=0.7\textwidth, angle=0]{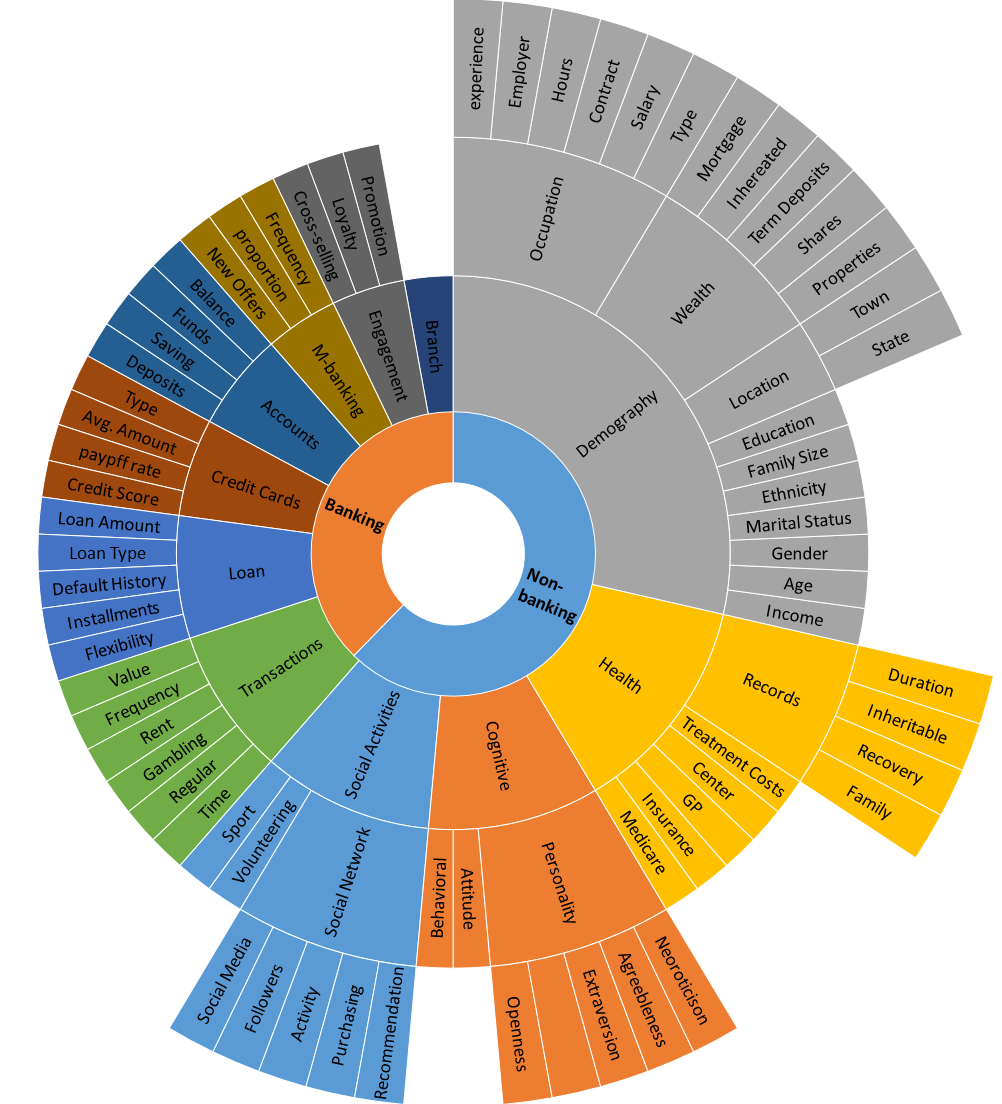}
    \caption{Customer Features Sunburst Chart.}
    \label{fig:sunburst-final}
\end{figure}

Regarding our model, we need a dataset that contains the banking and non-banking numeric features of customers which are discussed in Section 4.1. One of these features was the loan default history of the customer i.e., whether or not the customer has had any loan default history. We used this binary feature as the label of our classifier. The dataset we used for evaluation of our method is a Kaggle public dataset for loan-defaulter which contains more than 300k data with 122 attributes for each\footnote{https://www.kaggle.com/gauravduttakiit/loan-defaulter}. The dataset contains historical data about banking customers which includes their account and transaction details like average amount, frequency, transaction statement, and credit score in addition to some socio-demographic features such as age, income, and occupation information, AMR (Average Monthly Revenue), and AME (Average Monthly Expense). Regarding some features, we also used the data from one of the Australian banks by considering the privacy policies. To ensure privacy, profile names are masked and the data is in random order. For personality detection, we used texts of customers to extract personality scores. These texts are on external and social datasets such as customer reviews and feedback and also their text on social media. We matched the data of different datasets by mapping some common features including age, gender, education, job title, and location. For example, for customers of the same age group, gender, education, and occupation, we linked the features extracted from different datasets. The size of the dataset is 100000 observations and 63 dimensions for each of them. These dimensions include numeric banking features discussed in Section 4.1.1 such as loan details, transaction details, account details, and credit details. The dimensions of the dataset also include non-banking features discussed in Section 4.1.2 such as socio-demographic features and personality traits. The sunburst chart of all customer features used in the model is also shown in Figure~\ref{fig:sunburst-final}.

\subsection{Pre-Processing}

After extraction of all customer features and before training the model it is necessary to do some pre-processing activities in order to achieve the most accurate and trustful results. These pre-processing activities are generally divided into three main categories of analysis of personality features, standardization, and augmentation.

\subsubsection{Enabling the Analysis of Personality Features}

We leveraged state of the art approaches in extracting and detecting personality scores from the text as discussed in Section 4.1.2. Some of the text examples and 5 scores of personality traits that are extracted from this model are shown in Table ~\ref{fig:Personality Detection Samples.PNG}.

\begin{table}[ht]
 \caption{Extracting personality traits from customer texts.}
    \centering
    \includegraphics[scale=0.9]{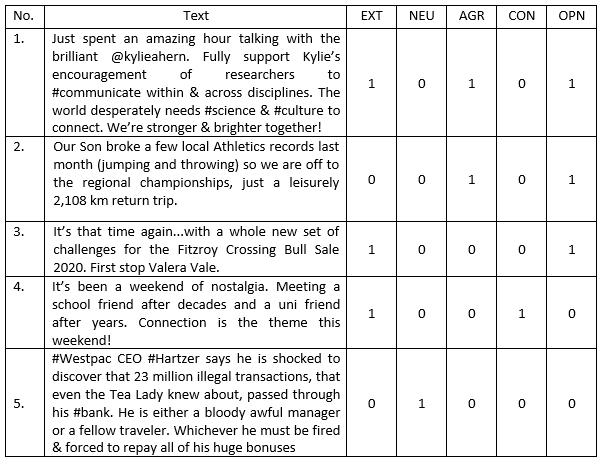}

    \label{fig:Personality Detection Samples.PNG}
\end{table}

\vspace{-0.5cm}

\subsubsection{Standardize the Input Data}

  As the ranges of different features are not similar and also the mean and variances are very different, we need to standardize them to make sure the model is not deviated by some of the attributes against the others. We did it by removing the mean and scaling to unit variance. We used the "sklearn.preprocessing.StandardScaler" package to standardize all of the numeric features.

 \subsubsection{Augmenting the Data}
  The dataset is skewed with less than 6 percent of positive labels which are the percentage of customers with a history of loan default. To achieve higher precision and recall, we need to augment the data to make the labels balanced and have more cases of positive labels. We used the "SMOTE" package for augmenting our data. Therefore, the shape of input datasets after applying "SMOTE" is augmented to 188170 rows of 63 features and a binary target.

 \subsection{Classification}

 Once personality traits detected and concatenated to the customer data and the data standardized and augmented, we used this data as the input vector into our fully connected neural network in which the first hidden layer was designed with 3 nodes and the second layer with 10 nodes. After training the model, the customer vector is embedded from the weights of the first hidden layer which causes the 3-dimensional vectors.

 \subsubsection{Neural Network Training}

 In order to predict the customer loan default probability and embed the customer vector, we trained our proposed neural network. To optimise the efficiency and getting the most accurate results, we tested different activation functions and measured the accuracy for each of them. We also tested different situations for the number of epochs and the batch size in splitting the data to find the best outcomes.

 \textbf{Activation Functions}.
 In the designed network there are 2 hidden layers and an output layer. We tried sigmoid function, tanh(x), Relu, and Leaky Relu for them and trained the model. The model had the best performance when we apply Leaky Relu for hidden layers and sigmoid for the output layer.

\begin{figure}[t]
    \centering
    \includegraphics[width=0.8\textwidth, angle=0]{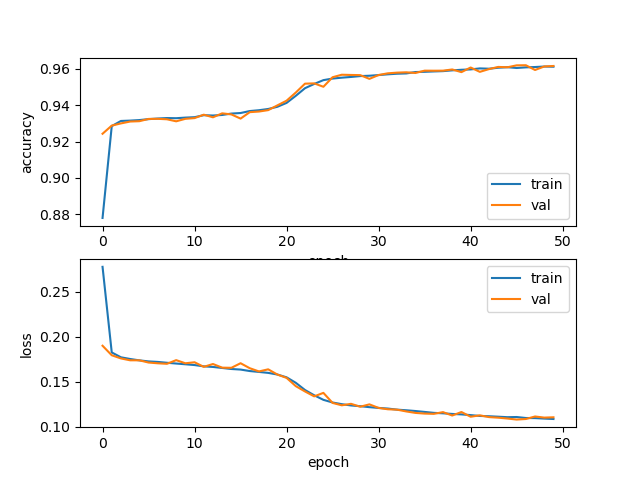}
    \caption{Accuracy and Loss in terms of number of epochs.}
    \label{fig:Acc_Loss_diagram.png}
\end{figure}

 \textbf{Data Split}.
 To perform the classification, we are dividing the data into 3 parts of 60 percent training data, 20 percent cross-validation, and 20 percent test data. The number of epochs and batch-size are both set to 50 and the optimization is based on the Adam model. We also set the early stop as 5 which means after 5 epochs with no improvement in loss decline, training will be stopped and weights will be finalised.
 Figure~\ref{fig:Acc_Loss_diagram.png} shows the increasing trend of accuracy and decreasing pattern of loss by raising the number of epochs.

\vspace{-0.5cm}

\subsubsection{Classification Evaluation}
 To evaluate the classification model we utilize, Accuracy score, MSE (Mean Squared Error), Precision, Recall, and F1 Score from the sklearn. The proportion of data with a positive label which is customers with a history of loan default are less than 6 percent within our dataset. Therefore, the "Accuracy" itself is not enough for the evaluation. For instance, if a model predicts all the results as 0 (predict all the customers to have no loan default), then the model will have more than 94 percent accuracy without an acceptable outcome. Therefore, in addition to accuracy, we are calculating precision, recall, and F1 score to evaluate our classification outputs. Below is the definition of these indexes:

  \begin{equation}\label{4.1}
    Accuracy = \frac{TP + TN}{(TP+TN+FP+FN)}
  \end{equation}\break

  \vspace{-1cm}

  \begin{equation}\label{4.2}
    mean-squared-error = \frac{1}{N}\sum\Vert Y_i -  Y_{predicted} \Vert ^2
  \end{equation}

\vspace{-1cm}

 \begin{multicols}{2}
  \begin{equation}\label{4.3}
    Precision = \frac{TP}{(TP+FP)}
  \end{equation}\break
  \begin{equation}\label{4.4}
    Recall = \frac{TP}{(TP+FN)}
  \end{equation}
\end{multicols}

 \vspace{-0.5cm}

  \begin{equation}\label{4.5}
    F-measure = \frac{2*Precision*Recall}{Precision+Recall}
  \end{equation}\break

  \vspace{-0.5cm}

In the above formulas, $Y_i$ stands for the real label of $i_{th}$ customer about the history of loan default, and $Y_{predicted}$ is the output of the classification model which is actually the output of the Sigmoid function of the last layer to predict whether or not the customer will have loan default.

Also, TP, TN, FP, and FN are representing true-positive, true-negative, false-positive, and false-negative respectively. True-positive are the cases that there was a loan default history for the customer and predicted correctly by the model. Similarly true-negative refers to the cases that the customer doesn't have any defaulted loan and the model also realizes this accurately. However false-positive cases are the customers that our model detects as customers that couldn't pay back their loan while they didn't have any default history and false-negative stands for customers with loan default history that the model is not able to identify.
The accuracy results of our experiment are shown in Table~\ref{table comparison}.

\begin{table}[ht]
\caption{Evaluation and accuracy of proposed classification method.}\label{table comparison}
\centering
 \begin{tabular}{||c c||}
 \hline
  & Classification with Proposed Method    \\
 \hline
 Accuracy  & 0.959  \\
 \hline
 MSE & 0.04\\
 \hline
  Loss  & 0.115\\
 \hline
 Precision  & 0.979 \\
 \hline
 Recall  & 0.942 \\
 \hline
 F-1 Score  & 0.961 \\
 \hline
\end{tabular}
\end{table}


\subsection{Clustering the Embedded Vectors}

Following the training of the fully connected neural network in the classification stage and finding the optimised weights, the 3-dimensional vectors are embedded from the hidden layer with 3 nodes. Then we aim to apply clustering on these embedded vectors to create a dashboard of customer segmentation in a 3-dimensional coordinate system.

\subsubsection{Clustering methods and packages}
\begin{figure}[t]
    \centering
    \includegraphics[width=0.5\textwidth, angle=0]{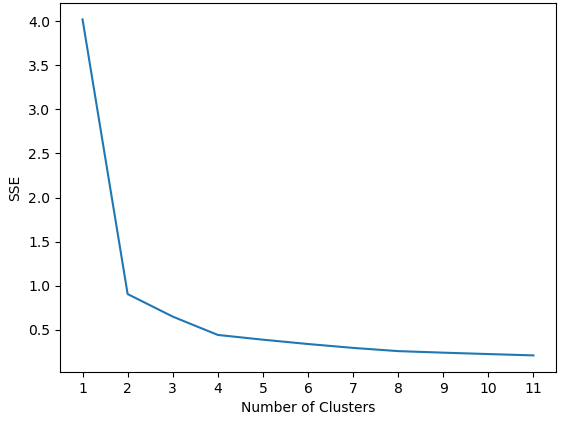}
    \caption{Knee method for finding the best number of clusters.}
    \label{fig:Kmeans knee graph.PNG}
\end{figure}

We use 4 different methods of clustering on the embedded vectors and compared the results. Kmeans, Mean-Shift, Gaussian Mixture, and SOM packages were used for clustering. While the mean-Shift model is able to find the most efficient number of clusters, for the other 3 models, we applied them with different numbers of clusters from 2 to 6 and compared the results. Regarding finding the most efficient k (number of clusters) we use the Knee model for k-means which is presenting the declining trend of SSE (Sum of Square Error) by increasing the number of clusters. Based on the Knee model, the best number of clusters is where the breakpoint is identified in the curve. However, there are some other considerations in choosing k such as business goals of segmentation and expert judgments. The knee graph is shown in Figure~\ref{fig:Kmeans knee graph.PNG} and we calculated the results for k=2,3,4,5,6.

\vspace{-0.4cm}

\subsubsection{Clustering Methods Comparison }

We apply four methods of clustering on our embedded vectors and compared the results of created clusters from these methods. We use 3 indexes to evaluate the cluster qualities which is comparing the homogeneity of objects inside a cluster with the minimum distance of 2 different clusters. The indexes are explained as below:

\begin{itemize}
    \item Silhouette-score: The best value is 1 and the worst value is -1. Values near 0 indicate overlapping clusters.
    \item Calinski and Harabasz(CH) score: higher Calinski-Harabasz score relates to a model with more efficient clusters.
    \item Davies-Bouldin Index: The minimum score is zero, with lower values indicating better clustering.
\end{itemize}

\subsubsection{Clustering Results}

\begin{figure}[t]
    \centering
    \includegraphics[width=0.8\textwidth, angle=0]{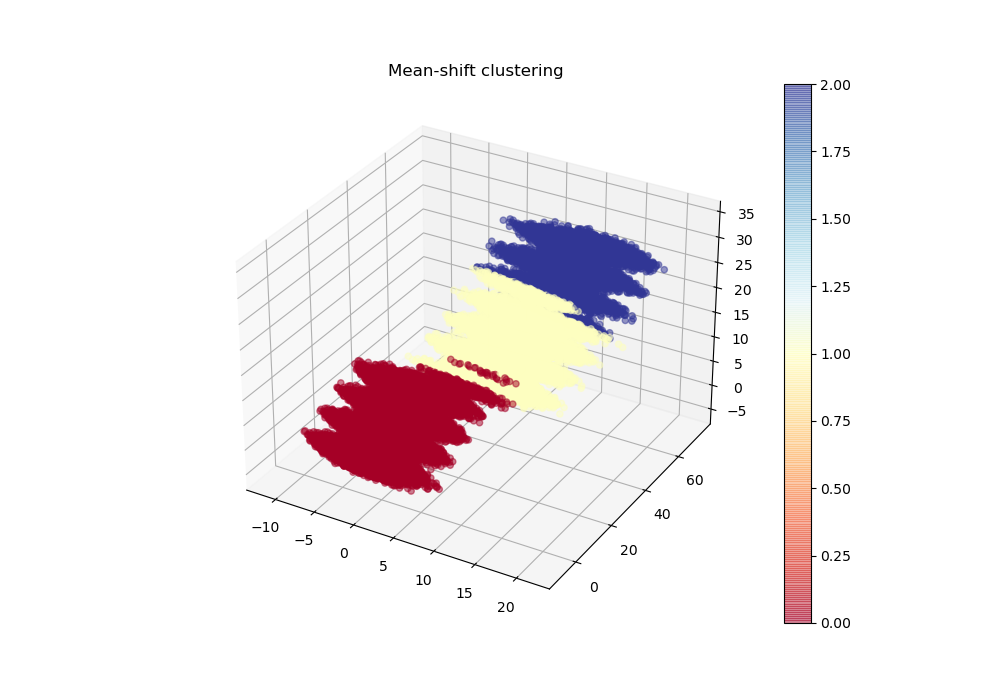}
    \caption{Mean-Shift embedded vector clustering.}
    \label{fig:Mean-Shift}
\end{figure}

We applied 4 methods of clustering and evaluated the results. The mean-Shift model is selecting the number of clusters itself based on the dataset which was 3 clusters in our experiment as can be seen in Figure~\ref{fig:Mean-Shift}.
The other 3 clustering methods i.e., k-means, Gaussian mixture, and SOM experimented with 2,3,4,5,and6 clusters for each, and the clustering results are shown in Table~\ref{fig:Clustering Evaluation}.

\begin{table}[ht]
\caption{Comparing results of clustering with different methods.}
\centering
\includegraphics[scale=0.9]{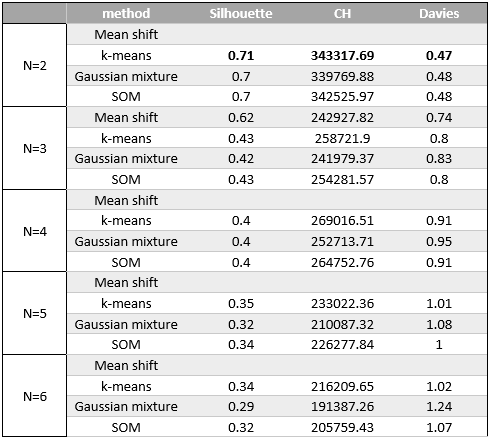}
\label{fig:Clustering Evaluation}
\end{table}

The results prove that the performance of different models with similar cluster numbers is comparable however, the k-means has slightly better results. In terms of scalability, all four methods were comparable and there wasn't a meaningful difference among them in the computation time by increasing the size of data. According to the number of clusters, based on the evaluation indices, the best results belonged to clustering with k=2,3 (2 or 3 clusters), however, the goal of the segmentation and the business needs should be considered in the number of clusters in the customer segmentation. For instance, banking experts might require more number of clusters and detailed segmentation due to their policies for customers. For example in our research, the banking analyst was interested in breaking the segmentation into smaller segments regarding different risk types and levels for more detailed investigation about the segment of risky customers.

The created clusters of the k-means model for embedded vectors regarding the different number of clusters from 2 to 6 are shown in Figure~\ref{fig:k-means clusters for k=2,3,4,5,6.PNG}.

\begin{figure}[t]
    \centering
    \includegraphics[width=1\textwidth, angle=0]{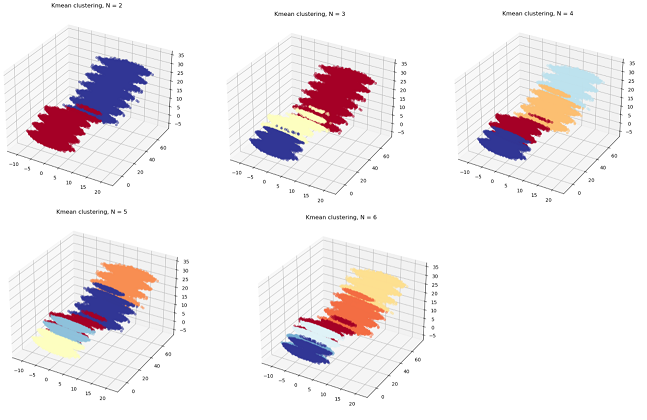}
    \caption{Embedded vectors clusters of k-means by k=2,3,4,5,6. }
    \label{fig:k-means clusters for k=2,3,4,5,6.PNG}
\end{figure}

\subsection{Discussion}

In this section, we reviewed the results of the proposed model of customer segmentation in terms of accuracy of classification, clustering, and embedded vector visualization.


\textbf{Subjective Customer Segmentation.}
The model proposed in this paper is able to perform subjective customer segmentation. Therefore, banking analyst is able to select the customer features based on the goal of segmentation and apply this model. The model provides the analyst with the segments following the banking attribute which is the goal of segmentation.

\textbf{Impact of Personality Traits}

One of the contributions of this paper is considering the personality traits as an input of the model. We extracted personality traits from customer texts and concatenate them into customer features. The results proved the impact of customer personality on their risk levels. Adding personality traits could rise the accuracy of loan default prediction.

\textbf{Similarity Detection}

The embedded vectors not only provides the analyst with a visualized customer segmentation but also improves the analytics and profiling methods by capturing customer similarities to a pre-defined customer and assist in the policies and services that will be offered to the customers. By embedding the vectors for each customer, banking analysts are able to measure the similarity of customer profiles whenever they need.


\section{Conclusion}

In this section, we summarize the contributions of this paper and discuss future research directions
to develop on this work.

\subsection{Concluding Remarks}
  Today, due to the extremely high level of competition in the market, proposing out of the box solutions for the existing customer requirements in addition to suggesting brilliant ideas about new services to satisfy the potential requirements are becoming the competitive advantage of companies. Thus, the significance of having true and clear insight into the customers is noticeable. Intelligent models and machine learning tools play the most critical role in this journey. Providing customers with the next best offers, maximizing the profit, and managing different types of risks requires novel intelligent methods of subjective and goal-based customer segmentation.

 In our research, we worked on an intelligent method of customer segmentation by embedding a vector for each customer and clustering these vectors. Therefore, the analyst will have the dynamic visualized customer segmentation and will be able to detect customers with similar behaviours and also will notice any changes in customer vectors over time.  We leveraged feature engineering to enable analysts to identify important features
(from a pool of features such as demographics, geography, psychographics, behavioral, and more) and feed them into the model.

\begin{figure}[t]
    \centering
    \includegraphics[width=0.6\textwidth, angle=0]{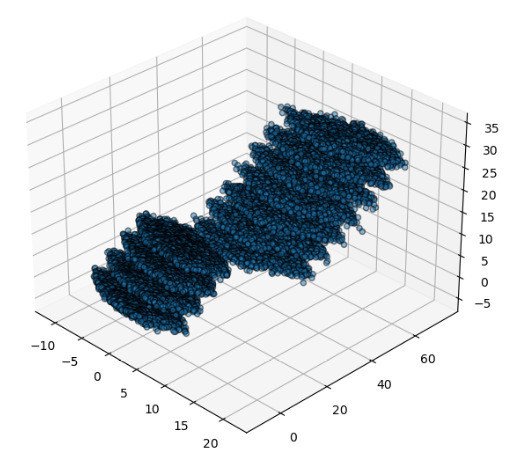}
    \caption{Customer Vectors in a coordinate system.}
    \label{fig:Customer Vectors in a coordinate system.png}
\end{figure}

 Our method contained both supervised and unsupervised methods in its pipeline. We used a fully connected neural network to classify the customers based on the goal of segmentation (customer credit risk in the motivating scenario of this research). This classification network also embedded a vector for each customer and the analyst will see their position in a 3-dimensional coordinate system as shown in figure~\ref{fig:Customer Vectors in a coordinate system.png}. Finally, we applied the clustering methods on the embedded vectors to identify customers with a high level of homogeneity within the clusters.

 \subsubsection{Customer2Vec Model Advantages in Profiling Systems}

 The model is able to assign a vector to each customer and by having this vector, apart from customer segmentation, banking experts will be able to calculate the similarity of customers. As a scenario, when the bank's customer is unsuccessful to repay the loan, he/she will be segmented as a high risk customer with regards to credit risk. Then bank expert will be able to immediately calculate the similarity of other existing customer vectors to this customer and re-segment them and/or providing them with specific services and apply some rules to detect the very similar customers and consider some policies in their existing requests~\cite{DBLP:conf/wise/TabebordbarBBB19}. Moreover, for a new application of loan from an existing or a new customer, the similarity of customer profile to the ones already had the loan default history will improve the accuracy of the profiling system.

 \subsection{Future Works}

 There are different areas that we can expand our proposed model. While this model is trying to result in a vector for each customer due to one specific attribute (credit risk in this research), it is possible to do this segmentation with multiple attributes. There are some techniques that we can improve our model with them. Below are the main domains that we plan to work on them in future works.

 \subsubsection{Applying Online Learning}
     As the behaviour of customers is changing over time, the online learning method which can be flexible due to the change of patterns can help us achieve great results. For example, regarding the credit risk prediction in the banking sector, while there are a huge amount of data of customers over time which is used to train the model, new causes of risks and new patterns of high risk customers will not be detected if we just use the old training data and try to predict credit risk of new customers. However, by applying online learning any new customer will also contribute to the training of the model which makes it flexible to discover behavioural pattern changes and conclude the updated weight of each feature in the target attribute. For example, the weight of parameters like age, income, and personality on the risk level of customer might change over time which online learning is possible to track these changes.

     \subsubsection{Dynamic Dashboard of Each Customer Over the Time}

 Over time features and attributes assigned to a customer will change. For example, age, income, spending patterns, etc, all are changing which will result in a change in his/her vector.
This model can be extended to provide a dynamic dashboard of all features during the time and some triggers to make the warnings for bank analysts in case of big changes in customer behavior. Using this visualized segmentation, it will be possible to have a view of high-risk customer behavior and vectors to make proactive decisions for a new customer being segmented in that part of an existing customer which changed in some features and moved to this area.

\subsubsection{Graph Model and OLAP}

By using the graph model to link two customer nodes based on one or some features in a specific banking domain attribute. So it will be possible to extend the customer2vec model to different attributes simultaneously and while two customers might be similar in an attribute, they might be not similar according to another attribute~\cite{Narayanan2017Graph2vec:Graphs}.
We also can extend the model in the future to reduce the dimension of the customer initial vector by using the PCA algorithm and/or OLAP warehousing to make it possible to project feature vector on some specific feature and focus on its impact on the model output.

\subsubsection{Crowd-Sourcing}

We will work on generating a database with all different required features and attributes of customers by applying crowd-sourcing~\cite{DBLP:conf/caise/BeheshtiVBT18} techniques to integrate different datasets and extract various objects and columns from them and link them together. By using more features from datasets and linking them together, we will improve the accuracy of the customer segmentation and similarity detection.


\section*{Acknowledgements}
- I acknowledge the AI-enabled Processes (AIP\footnote{https://aip-research-center.github.io/}) Research Centre and Tata Consultancy Services for funding this project.

\bibliographystyle{abbrv}
\bibliography{ms}

\end{document}